\documentclass[11pt]{article}
\pdfoutput=1
\usepackage{jheppub}
\usepackage{graphicx}
\usepackage{amssymb}
\usepackage{bbm}
\usepackage{mathrsfs}
\usepackage{amsmath,amssymb}
\usepackage{slashed}
\usepackage{hyperref}
\usepackage{caption}
\usepackage{xcolor}
\usepackage{dsfont}
\usepackage{verbatim}
\usepackage{subfig}
\usepackage{amsmath,amssymb,amscd,amsfonts,mathtools}
\usepackage{xifthen}
\usepackage{xcolor}
\definecolor{markgreen}{RGB}{230,243,230}
\definecolor{darkolivegreen}{rgb}{0.33, 0.42, 0.18}
\definecolor{darkpastelgreen}{rgb}{0.01, 0.75, 0.24}

\usepackage[toc,page]{appendix}
\usepackage{epsfig}
\usepackage{epstopdf}
\usepackage{latexsym}
\usepackage{graphicx}
\usepackage{booktabs}
\usepackage{bbm}
\usepackage{color}
\usepackage{physics}
\usepackage{tensor}
\usepackage{verbatim}
\usepackage{subfig}
\usepackage{tikz}
\usepackage{ifthen}
\usepackage{array}
\usepackage{csquotes}
\usetikzlibrary{matrix}
\usetikzlibrary{decorations.markings,calc,shapes,decorations.pathmorphing,patterns,decorations.pathreplacing,arrows.meta}
\usetikzlibrary{positioning}
\usepackage{hyperref}
\usetikzlibrary{patterns}
\usetikzlibrary{positioning}
\newdimen\mydim
\pdfoutput=1
\makeatletter
\def\@fpheader{\relax}

\newcommand*{\ov}[1]{
  $\m@th\overline{\mbox{#1}}$
}
\newcommand*{\ovA}[1]{
  $\m@th\overline{\mbox{#1}\raisebox{3mm}{}}$
}
\newcommand*{\ovB}[1]{
  $\m@th\overline{\mbox{#1\rule{0pt}{3mm}}}$
}
\newcommand*{\ovC}[1]{
  $\m@th\overline{\mbox{#1\strut}}$
}
\newcommand*{\ovD}[1]{
  $\m@th\overline{\mbox{#1\vphantom{\"A}}}$
}
\newcommand*{\ovE}[1]{
  $\m@th\overline{\raisebox{0pt}[1.2\height]{#1}}$
}
\newcommand*{\ovF}[1]{
  $\m@th\overline{\raisebox{0pt}[\dimexpr\height+1mm\relax]{#1}}$
   Package `calc' can be used as alternative for `\dimexpr'.
}
\newcommand*{\ovG}[1]{
  $\m@th\overline{\raisebox{0pt}[\dimexpr\height+1mm\relax]{#1\vphantom{A}}}$
}
\makeatother
\newboolean{showcomments}
\setboolean{showcomments}{false}
\newcommand\rem[1]{\ifthenelse{\boolean{showcomments}}{{#1}}{}}
\newcommand{\be}{\begin{equation}}
\newcommand{\ee}{\end{equation}}

\usepackage{xifthen}
\newcommand{\dalembert}[1][]{\ifthenelse{\isempty{#1}}{\Box}{#1\Box}}
\usetikzlibrary{decorations.pathmorphing}
\tikzset{snake it/.style={decorate, decoration=snake}}

\title{\Large Analyzing the general conditions for modulus stabilization in a warped braneworld}
\author{Soham Bhattacharyya$^{a}$ and Soumitra SenGupta$^{a}$}
\affiliation{$^{a}$School of Physical Sciences, Indian Association for the Cultivation of Science, Kolkata-700032, India.}
\emailAdd{soham1316@gmail.com, tpssg@iacs.res.in}

\abstract{In braneworld scenarios with compact extra dimensions, the modulus field typically remains undetermined without an appropriate stabilization mechanism. A common approach introduces a bulk scalar field that generates an effective potential for the modulus with a stable minimum. In this work, we explore some novel aspects of such stabilization mechanisms. We study how the bulk scalar profile influences the stabilization procedure. Following the approach of Chacko et al.
\cite{Chacko1}, we analyze several representative cases using methods of singular perturbation theory. We identify a consistent relationship between the structure of the bulk potential and the emergence of a stabilized modulus, and outline the general conditions that any bulk potential must satisfy to enable stabilization. In this context, we also examine a potential connection between geometric consistency conditions—specifically, the \enquote{brane world sum rules}—and the stabilized value of the modulus. In some scenarios where stabilization occurs, we find that these sum rules can offer additional constraints on the modulus, providing a complementary perspective on its determination. Taken together, these results offer a broader perspective on the mechanisms that govern modulus stabilization in higher-dimensional warped geometries.}

\begin{document}
\maketitle
\flushbottom
\pagebreak

\section{Introduction}\label{sec1}
\noindent The gauge hierarchy problem in Standard Model (SM) of particle physics results into the well known fine-tuning problem in connection to the Higgs mass which acquires a quadratic divergence due to
the large radiative corrections in perturbation theory.
In order to confine the Higgs mass parameter within TeV scale, one needs to consider theories beyond SM.
Among many such attempts, the two-brane Randall-Sundrum (RS) 
warped extra-dimensional scenario
has earned special attention for 
following reasons \cite{RS1}: (a) it potentially resolves the gauge hierarchy problem 
without introducing any other intermediate scale in the theory and (b) the extra-dimensional modulus can be stabilized by introducing
a bulk scalar field in the setup \cite{GW1}.\\

\noindent One of the crucial aspects of such braneworld models is to stabilize the distance between the two
branes (known as modulus or radion). For this, one needs to generate an appropriate 
potential for the radion field with a stable minimum consistent with the value proposed in RS model
in order to solve the gauge hierarchy problem.
Goldberger and Wise (GW) \cite{GW1,GW2} proposed a very useful mechanism to achieve this by 
introducing a bulk scalar field with suitable bulk and brane potentials. They showed that one can indeed
stabilize the modulus without any unnatural fine-tuning of the parameters, though ignoring effects of the backreaction 
of the bulk scalar on the background metric.\\

\noindent Since the GW proposal, different stabilization procedures have been proposed and among these, DeWolfe et al. \cite{Freedman} is one of the most prominent ones as it provided an exact solve including the backreaction, which was missing in the original GW prescription. Later on, Chacko et al. \cite{Chacko1,Chacko2} revisited the GW stabilization including the gravitational contribution to radion potential and also including self interaction terms for the bulk potential, albeit neglecting backreaction. They heavily relied on an approximate solution for the scalar field profile (for a non-trivial bulk potential) and this is something we have discussed in some detail here as well. Several variants of the RS model, their respective modulus stabilization and cosmological implications were explored previously in Refs. \cite{csaki2001,gwvariations,geometry_modulus,SauryaDas_2008,PhysRevD.75.107901,Das_mukherjee,tanmoy2,banerjeepaul,Banerjeessg, tanmoy,inflation_ssg, Randall_2023,ashmitassg,Koley_2009}. Quantum effects in such theory were studied in Refs. \cite{odinstov1,odinstov2}.\\

\noindent Given all this research, an evident question to ask is how does the scalar field profile affect the modulus stabilization procedure? Is there a correspondence between the nature of the GW bulk potential and successful modulus stabilization? This is the problem we set out to explore in this work. Apparently, the relationship between the two seems highly non-trivial because of the sheer number of steps involved in reaching upto the radion potential starting from the scalar field profile. So, the way we wish to tackle this problem is by exploring the modulus stabilization scheme taking different scalar actions, incorporating different forms for the bulk potentials. It must be noted that in AdS space which is the nature of the bulk in Randall-Sundrum, the stability of the GW field only requires satisfying the Breitenlohner-Freedman (BF) bound \cite{1982}: $m^2/k^2+4>0$ where $m^2$ is the mass squared of the GW field and $k$ is the AdS curvature scale. Thus, though the presence of a minimum in the GW bulk potential clearly indicates a stable field; the lack of a minimum does not mean an unstable GW field as long as the BF bound is satisfied.\\

\noindent The primary scenarios that we discuss and provide a detailed analysis for are the following: (a) stabilizing scalar with no bulk interactions, (b) bulk cubic interaction, (c) bulk quartic interaction, (d) bulk double-well potential, (e) Bazeia-Furtado-Gomes (BFG) type bulk potentials, and (e) Mishra-Randall ansatz bulk potentials. We highlight a correspondence between the nature of the bulk scalar potential and modulus stabilization. Lastly, we provide a simplified criterion that any bulk scalar potential must satisfy to enable modulus stabilization.\\

\noindent The other aspect of the paper is the following: it is a known fact that checking for modulus stabilization and thereby getting to the stabilized modulus value is cumbersome for any non-trivial bulk potential - we solve for the scalar field profile - plug it back into the GW action - integrate over the extra dimension - add gravitational contributions (if necessary) to it - and work up to the leading order to get the radion potential. So, a natural question to ask is whether there exists an alternative approach that could yield the stabilized modulus value (if it gets stabilized), at least in certain cases? Here we show that one could indeed avoid all the hassle and still get a bound on the stabilized modulus ($r_c$), employing the brane world sum rules, put down by Gibbons et al. in Ref. \cite{Gibbons} in certain scenarios. We also show that these bounds are actually close to the true stabilized modulus values that one gets by minimizing the radion potential. This is especially interesting as it might serve as a go-to prescription for estimating the stabilized modulus value and a hint at possibly unexplored physics.\\

\noindent In the context of brane world sum rules, we must mention that their extensions and applications \cite{Leblond,abdalla2009notes,hoff2011five,hoff2010torsion,abdalla2010positive,fabris2018braneworld,de2015smooth,peyravi2023thick,hoff2011two,da2017f,dias2015thick,abdalla2009consistency} have been thoroughly studied by the community over the past two decades. An immediate extension to higher-dimensional braneworlds was presented in Ref. \cite{Leblond}. Since then, the consistency conditions have been used as tools to probe the existence of thick branes and positive tension branes in several modified theories of gravity - f(R) \cite{hoff2011five}, non-conservative \cite{fabris2018braneworld}, higher-order \cite{peyravi2023thick}, Einstein-Palatini f(R) \cite{da2017f}, Einstein-Gauss-Bonnet \cite{dias2015thick}, Brans-Dicke \cite{abdalla2010positive}, scalar-tensor theories \cite{abdalla2009consistency} and torsional gravity \cite{hoff2010torsion}. These rules have also been used to study the two-brane variable tension model - a particularly significant result being, if the hidden brane tension obeys \textit{E$\ddot{o}$tv$\ddot{o}$s} law, then the corresponding visible brane tension shows a bouncing behaviour \cite{abdalla2009notes,hoff2011two}.\\

\noindent The structure of the paper is as follows. Sec.\ref{sec2} provides a quick review of the RS-I setup and the original GW prescription. Sec.\ref{sec3} discusses a more general form of the GW scheme involving gravitational contributions to radion potential and bulk interaction terms for the GW scalar. Sec.\ref{sec4} deals with the six different scenarios we mentioned previously \footnote{It is to be noted that when we talk of radion stabilization, we always take the radion potential generated only due to the GW part and not the gravitational sector contribution, unless mentioned otherwise (Sec.\ref{sec7} includes the gravitational contribution).}. Sec.\ref{sec5} highlights the correspondence between the nature of bulk scalar potentials and successful radion stabilization. Sec.\ref{sec6} provides the criterion on any bulk scalar potential to achieve modulus stabilization. Sec.\ref{sec7} discusses the connection with brane world sum rules. A short conclusion is provided in Sec.\ref{sec8}.
\section{Warped extra-dimensions and modulus stabilization}\label{sec2}
\subsection{Randall-Sundrum I}\label{sec2.1}
The geometry is nonfactorizable with following form of metric tensor:
    \begin{equation}\label{1}
    \begin{split}
        ds^2 = G_{MN}dx^Mdx^N = e^{- 2 \sigma} \eta_{\mu \nu} dx^{\mu} dx^{\nu} - r_c^2 d \phi^2
    \end{split}
    \end{equation}
    where $\sigma$ is the warp factor, the greek index runs from 0 to 3 while the latin index runs from 0 to 5, excluding 4 with 5 signifying the fifth spatial dimension, and $r_c$ is the radius of compactification of the extra dimension. We work with the mostly negative metric signature. The topology of the extra dimension is $S^1/\mathbb{Z}_2$. The fifth coordinate is angular in nature with $(x,\phi)$ identified with $(x,-\phi)$. The 3-branes extending in $x^\mu$ direction are located at $\phi=0$ and $\phi=\pi$. The induced metrics on the branes are given by:
    \begin{equation}\label{2}
        g_{\mu \nu}^{vis}(x^{\mu}) \equiv G_{\mu \nu}(x^{\mu}, \phi = \pi), \ \  \ 
        g_{\mu \nu}^{hid}(x^{\mu}) \equiv G_{\mu \nu}(x^{\mu}, \phi = 0)
    \end{equation}          
The classical action for the setup is given by:
\begin{eqnarray}\label{3}
S &=& S_{gravity} + S_{vis} + S_{hid} \nonumber \\
S_{gravity} &=& 
\int d^4 x \int_{- \pi}^{\pi} d \phi \sqrt{-G} \{- \Lambda_b - 2 M^3 R \} 
\nonumber \\
S_{vis} &=& \int d^4 x \sqrt{-g_{vis}} \{ {\cal L}_{vis} 
-  T_{vis} \} \nonumber \\
S_{hid} &=& \int d^4 x \sqrt{-g_{hid}} \{ {\cal L}_{hid} -  T_{hid} \}
\end{eqnarray}
where M is the 5D Planck mass. It is to be noted that $T_{vis}$ and $T_{hid}$ are vacuum energies of the branes which act as gravitational sources even in the absence of matter on the branes. Thus, without putting matter on the branes, we obtain the  following equation by varying the action with respect to the bulk metric:

\begin{equation}\label{4}
\begin{split}
\sqrt{-G} ( R_{MN}-{1 \over 2 } G_{MN} R) = - \frac{1}{4 M^3} 
[ \Lambda \sqrt{-G} G_{MN} +  T_{vis}\\ \sqrt{-g_{vis}} g_{\mu \nu}^{vis} \delta^\mu_M \delta^\nu_N \delta(\phi - \pi)+
 T_{hid} \sqrt{-g_{hid}}  g_{\mu \nu}^{hid} 
\delta^\mu_M \delta^\nu_N \delta(\phi) ]
\end{split}
\end{equation}
The solution satisfying four dimensional Poincare invariance in the $x^\mu$ direction and respecting the orbifold symmtery is given by
\begin{equation}\label{5}
    \sigma =  r_c |\phi| \sqrt{\frac{- {\Lambda_b}}{24M^3}}
\end{equation}
with the following consistency criteria for the brane tensions
\begin{equation}\label{6}
    T_{hid} = - T_{vis} = 24 M^3 k, ~~ \Lambda_b = - 24 M^3 k^2
\end{equation}
where $k = \sqrt{\frac{- {\Lambda_b}}{24M^3}}$. From Eq.(\ref{5}), we get that $\Lambda_b$ has to be negative resulting in the bulk spacetime between the branes being a slice of an $AdS_5$ geometry. Hence, our solution for the bulk metric is then given by
\begin{equation}\label{7}
    ds^2 = e^{- 2 k r_c |\phi|} \eta_{\mu \nu} dx^{\mu} dx^{\nu} - r_c^2 d \phi^2
\end{equation}
\subsection{Why modulus stabilization?}\label{sec2.2}
\noindent The 5D RS action without any bulk scalar is given by Eq.(\ref{3}). Starting from this 5D action involving gravity only, the effective 4D action involving the 4D graviton and radion fields is found out by plugging in the 5D Ricci scalar $\mathcal{R}$ and integrating over the extra dimension. Consequently, the RS metric that we use is (we promote the metric and modulus to be dynamical fields)
\begin{equation}\label{8}
    ds^2 = e^{-2kr(x)|\phi|}g_{\mu\nu}dx^\mu dx^\nu - r^2(x)d\phi^2
\end{equation}
In absence of the 4D graviton, $g_{\mu\nu} \rightarrow \eta_{\mu\nu}$ as in the original RS prescription. Defining $\varphi\equiv Ae^{-k\pi r(x)}$, where $A=\sqrt{24M^3/k}$ to be the canonical radion field, one obtains the following form for the gravitational contribution to radion potential \cite{Chacko1}
\begin{equation}\label{9}
\begin{split}
    V_{GR}(\varphi)= \frac{k^4}{A^4}\left(\tau\:\varphi^4+\Lambda_{4D}\right)
\end{split}
\end{equation}
where $\Lambda_{4D}=\Big(T_{hid}+\frac{\Lambda_b}{k}\Big)/k^4$ is the $4D$ cosmological constant which we tune to be small, and $\tau=\Big(T_{vis}-\frac{\Lambda_b}{k}\Big)/k^4$. Thus, the potential is essentially $V_{GR}(\varphi)= \tau (k\varphi/A)^4$. Originally, \cite{RS1} set this potential to zero through fine tunings of the brane tensions i.e. $T_{vis} = -T_{hid} = \Lambda_b/k$ (see Eq.\ref{6}). However, then the radion field could take up any possible value. To fix this, one needed a stabilization mechanism that fixes the interbrane separation and makes the radion acquire a mass. This is essential as the stabilized modulus enters as a parameter in the lower dimensional effective theory.

\subsection{Goldberger-Wise stabilization}\label{sec2.3}
\noindent In a realistic RS scenario, a mechanism is needed to stabilize the 
geometry and give the radion a mass. Such a mechanism was proposed by 
Goldberger and Wise \cite{GW1}. In this construction, a 
massive $5D$ field $\Phi$ is sourced at the boundaries and acquires a 
VEV whose value depends on the location in the extra dimension. After 
integrating over the extra dimension, this generates a radion potential in the low energy effective theory. In the original GW 
construction, the quartic potential for the radion from the gravity sector was 
tuned to zero, and only the dynamics of the scalar field $\Phi$ 
contributed to the radion potential.\\

\noindent We add to the original RS action a scalar field $\Phi$ with the following bulk action
\begin{equation}\label{10}
S_b={1\over 2}\int d^4 x\int_{-\pi}^\pi d\phi \sqrt{G} \left(G^{AB}\partial_A \Phi \partial_B \Phi - m^2 \Phi^2\right)
\end{equation}
where $G_{AB}$ with $A,B=\mu,\phi$ is given by Eq.(\ref{7}).  We also include interaction terms on the hidden and visible branes (at $\phi=0$ and $\phi=\pi$ respectively) given by
\begin{equation}\label{11}
S_h = -\int d^4 x \sqrt{-g_h}\lambda_h \left(\Phi^2 - v_h^2\right)^2
\end{equation}
and
\begin{equation}\label{12}
S_v = -\int d^4 x \sqrt{-g_v}\lambda_v \left(\Phi^2 - v_v^2\right)^2
\end{equation}
We get a $\phi$-dependent vacuum expectation value $\Phi(\phi)$ which is determined classically by
\begin{equation}\label{14}
\Phi(\phi) = e^{2\sigma}[A e^{\nu\sigma}+B e^{-\nu\sigma}]
\end{equation}
where $\sigma(\phi)= kr_c |\phi|$ with $\nu=\sqrt{4+m^2/k^2}$.  Plugging this solution back into the scalar field action and integrating
over $\phi$ yields an effective four-dimensional potential $V_\Phi$ for $r_c$ which involves the constants A and B. These are determined from the boundary conditions i.e. matching with the delta functions. In the limit of large $\lambda_h$, $\lambda_v$ and $kr_c$, $\Phi(0)=v_h$ and $\Phi(\pi)=v_v$ are the energetically favoured boundary conditions. In the large $kr_c$ limit, the radion potential turns out to be
\begin{equation}\label{17}
\begin{split}
V_{GW}(r_c)= k\epsilon v_h^2 + 4ke^{-4kr_c\pi}(v_v - v_h e^{-\epsilon kr_c\pi})^2\left(1+\frac{\epsilon}{4}\right) - k\epsilon v_h e^{-(4+\epsilon)kr_c\pi}(2 v_v - v_h e^{-\epsilon kr_c\pi})
\end{split}
\end{equation}
where $m/k\ll 1$ so that $\nu=2+\epsilon,$ with $\epsilon \approx m^2/4k^2$ a small quantity. Terms of order $\epsilon^2$ are neglected though $\epsilon kr_c$ is not neglected. Ignoring terms proportional to $\epsilon$, this potential has a minimum at 
\begin{equation}\label{18}
kr_c = \left(\frac{4}{\pi}\right) \frac{k^2}{m^2} \ln\left[\frac{v_h}{v_v}\right]
\end{equation}
Using $v_h/v_v=1.5$ and $m/k = 0.2$ in Eq.(\ref{18}) yields $kr_c\approx 12$ and as one can see, no unnatural fine-tuning of parameters is required to solve the hierarchy problem.
\section{Generalized Goldberger-Wise stabilization}\label{sec3}
\noindent In the original Goldberger-Wise (GW) construction, the quartic potential for the radion $V_{GR}(\varphi)$ was fine-tuned to zero, leaving the dynamics of the scalar field $\Phi$ as the sole contributor to the radion potential. However, it is possible to relax this constraint by allowing a non-zero quartic contribution from the gravitational potential and incorporating self-interaction terms in the bulk potential of the GW scalar. This generalized approach has been effectively explored by Chacko, Mishra, and Stolarski in Ref.\cite{Chacko1}. We henceforth call this approach the CMS method, after the authors of Ref.\cite{Chacko1}.
\subsection{Approximate solution for the scalar field profile}\label{sec3.1}

\noindent The action for the GW scalar is given by
\begin{equation}\label{19}
\begin{split}
\mathcal{S}_{GW}
=\int d^4x\:d\phi\:
[
\sqrt{G}\:(
\frac12G^{AB}\partial_A\Phi\partial_B\Phi
-V_b(\Phi)
)
-\sum_{i=\{vis,hid\}}\delta(\phi-\phi_i)\:\sqrt{-G_i}\:V_i(\Phi)
] \; 
\end{split}
\end{equation}
where $\Phi$ is the GW scalar, $V_b(\Phi)$ is the bulk scalar potential and $V_{hid}(\Phi)$ ($V_{vis}(\Phi)$) is the brane potential on the hidden (visible) brane. In the Chacko et al. parametrization, the radion does not couple to the 
hidden brane at $\phi_h = 0$. Hence, we need not specify any form of 
$V_{hid}$, but only require that it sets $\Phi$ at $\phi=0$ to be 
$k^{3/2}v$ and that it does not contribute to the hidden brane tension. One such choice for the hidden brane potential is $\gamma (\Phi^2 - k^3v^2)^2$.
On the visible brane, they consider a linear potential for $\Phi$ of the form
 \begin{eqnarray}\label{20}
V_{vis}(\Phi)&=&2k^{5/2}\alpha\:\Phi \; 
 \end{eqnarray}
 where $\alpha$ is a dimensionless number. As is known and Ref.\cite{Chacko1} states as well, the choice of a linear potential for $\Phi$ on the visible brane is natural and is generally expected when $\Phi$ is not charged under any symmetries. For example, if $\Phi$ were charged under a $\mathbb{Z}_2$ symmetry i.e. $\Phi \rightarrow -\Phi$, then the linear term would be forbidden, and the leading allowed term would be quadratic or higher.\\

\noindent The bulk potential for the GW scalar $\Phi$ has the general form
 \begin{eqnarray}\label{21}
V_b(\Phi) = \frac12m^2\Phi^2+\frac{1}{3!}\eta\Phi^3+...\;\;\;
 \end{eqnarray}	
Given the action, we can solve for the scalar field profile $\Phi(\phi)$ in the RS-I background. The equation satisfied by $\Phi$ in 
the bulk with the boundary conditions resulting from brane potentials is 
given by
 \begin{eqnarray}\label{22}
&&\partial_\phi^2\Phi
-4kr_c\partial_\phi \Phi
-r_c^2V_b'(\Phi)=0
\nonumber\\
&&\phi=0\qquad:\qquad \Phi=k^{3/2} v
\nonumber\\
&&\phi=\pi\qquad:\qquad \partial_\phi\Phi=-\alpha k^{3/2} kr_c\;\;\
 \end{eqnarray}
We consider solutions for $0 \leq \phi \leq \pi$ because of the ${\mathbb{Z}}_2$ orbifold symmetry.
It follows that in the limit of large $4kr_c$, two independent approximate solutions to Eq.(\ref{22}) can be obtained in the following way: once 
by dropping the $\partial_\phi^2\Phi$ term, and once by dropping the potential term. This is motivated by methods of singular perturbation theory. These two 
equations are denoted as the outer region (OR) and boundary region (BR) 
equations respectively. The OR solution holds in the bulk, 
while the BR solution holds close to the visible brane where a boundary layer 
is formed with thickness of the order $\sim 1/4kr_c$.
\begin{equation}\label{23}
 \frac{d\Phi}{d\phi} = -\frac{r_c}{4k}V_b'(\Phi) \; \; \; \; \; \; \; \; \textbf{(OR)}
\nonumber
\end{equation}
\begin{equation}
\; \; \frac{d^2\Phi}{d\phi^2} = 4kr_c\frac{d\Phi}{d\phi} \; \; \; \; \; \; \; \;   \textbf{(BR)}
\end{equation}
The BR solution being independent of the bulk potential $V_b$ is readily solved. On applying the 
boundary condition (b.c.) at $\phi = \pi$, the BR solution is given by
\begin{eqnarray}\label{24}
\Phi_{BR}(\phi)= -\frac{k^{3/2}\alpha}{4}e^{4kr_c(\phi-\pi)} +C\;  
\end{eqnarray}
The constant $C$ is determined by asymptotic 
matching to the OR solution \cite{Bender2010-wx}. Hence, we get a smooth solution that very well approximates the two independent solutions in their regions of relevance. The parameters $\alpha$ and $v$ are chosen to be small and almost of the same order.

\subsection{Getting the stabilized modulus}\label{sec3.2}
\noindent On getting the approximate scalar field solution, we plug it back into the GW action (Eq.\ref{19}) along with the corresponding bulk and brane potentials. We then perform dimensional reduction, add gravitational contributions to it i.e. $V_{GR}(\varphi)$ and work up to the leading order to get the radion potential. Effectively, in presence of the GW scalar, the $\tau$ in $V_{GR}(\varphi)$ will receive additional corrections. At the minimum of this radion potential, we say that the modulus is stabilized and the corresponding value of $r_c$ is what we seek (see Appendix B, \cite{Chacko1} for details).

\section{In search of a stabilized modulus}\label{sec4}
\subsection{No bulk interactions}\label{sec4.1}
\noindent In the Goldberger-Wise prescription, they used a stabilizing scalar with no bulk interaction. But now suppose that we flip the sign of the kinetic term (phantom instability) or the mass term (tachyon instability) in Eq. (\ref{10}). We wish to now know whether we still find a stabilized modulus for our model. We should follow same steps as previously and write down the new classical equation of motion for the bulk scalar,
\begin{equation}\label{25}
    \partial^2_\phi \Phi - 4\frac{\sigma}{\phi}\partial_\phi \Phi + m^2 r_c^2 \Phi = 0
\end{equation}
where $\sigma = kr_c|\phi|$. The general solution to Eq.(\ref{25}) is given by
\begin{equation}\label{26}
    \Phi = e^{2\sigma}[Ae^{\gamma \sigma} + Be^{-\gamma \sigma}]
\end{equation}
where $\gamma = \sqrt{4-\frac{m^2}{k^2}}$. Thus, to get a real scalar field (avoiding oscillatory solutions), we need to have $m^2 \leq 4k^2$. This is well in agreement with the BF bound \cite{1982} which says that in AdS space, the mass squared parameter $m^2$ for the GW 
scalar $\Phi$ can be negative without giving rise to instabilities as 
long as the condition $m^2/k^2+4>0$ is 
satisfied.\\\\
The intermediate equations (as in Ref.\cite{GW1}) get reproduced in the same manner except the $\nu$ getting replaced by $\gamma$ in all the expressions.\\

\noindent Taking $m/k << 1$ so that $\gamma = 2-\epsilon$ with $\epsilon = m^2/4k^2$, the potential takes the form
\begin{equation}\label{27}
\begin{split}
    V_{GW}(r_c)= -k\epsilon v_h^2 + 4ke^{-4kr_c\pi}(v_v - v_h e^{\epsilon kr_c\pi})^2\left(1-\frac{\epsilon}{4}\right)
    + k\epsilon v_h e^{-(4-\epsilon)kr_c\pi}(2 v_v - v_h e^{\epsilon kr_c\pi})
\end{split}
\end{equation}
Ignoring terms proportional to $\epsilon$, the potential has a minimum at 
\begin{equation}\label{28}
kr_c = \left(\frac{4}{\pi}\right) \frac{k^2}{m^2} \ln\left[\frac{v_v}{v_h}\right]
\end{equation}
Thus, we see that if $v_v$ and $v_h$ retain their previously assigned values (i.e. keeping the brane potentials fixed) which made the original GW scenario stable, we no longer have a modulus stabilization. The reason for this is that we had $v_v<v_h$ in the original setup which renders $kr_c$ to be negative (unphysical) from Eq.(\ref{28}). What it essentially means is that we get a monotonic radion potential in the positive $r_c$ sector. Hence, we see that given fixed brane potentials, flipping sign of either term in the GW action results in modulus destabilization. An obvious question is - what if both the terms were made to flip signs? That would result in a violation of the weak energy condition (WEC) for the energy-momentum (EM) tensor of the scalar field (the original EM tensor for the scalar field was assumed to satisfy the WEC - so the new EM tensor which is just the negative of the previous no longer satisfies WEC\footnote{WEC states that $T_{\mu\nu}u^\mu u^\nu \geq 0$ where $T_{\mu\nu}$ is the energy-momentum tensor and u is any timelike vector.}). It is worth mentioning now that for each of the cases that we discuss here, when we discuss the inverted potential, we would be working with the same brane potentials and boundary conditions. This is essential to ensure a consistent comparison. In other words, we could have chosen the brane potentials such that we get a minimum in the inverted case but under the same conditions, we would not get an extremum in the original GW case. The equivalent analysis for CMS boundary conditions can be found in Sec.\ref{sec5}.\\

\noindent Here, we comment that whenever we flip the sign of the potential term keeping a canonical kinetic term, it is equivalent to having the original potential and a phantom kinetic term. This is because the bulk scalar equation would be identical for either case as is evident from Eq.(\ref{19}).
\subsection{Bulk cubic interaction}\label{sec4.2}
\noindent Consider a cubic self interaction term in the bulk potential.
\begin{eqnarray}\label{29}
V_b(\Phi)=\frac12m^2\Phi^2+\frac{1}{3!}\eta\,\Phi^3 \; 
\end{eqnarray}
We now employ the CMS scheme. In the limit where the cubic dominates the potential, the OR solution is given by
 \begin{eqnarray} \label{30}
\Phi_{OR}(\phi)= \frac{k^{3/2}v}{1+\xi kr_c\phi} \; 
 \end{eqnarray}
where $\xi = \eta v/8\sqrt{k}$ and we have imposed the boundary 
condition $\Phi(0)=k^{3/2}v$. Combining the OR and the BR solutions, we obtain the 
complete solution for $\Phi$ as
 \begin{eqnarray} \label{31}
\Phi_{approx}(\phi)=-\frac{k^{3/2}\alpha}{4}e^{4kr_c(\phi-\pi)}
+\frac{k^{3/2}v}{1+\xi kr_c\phi} \; 
 \end{eqnarray}
Note that this solution is ill-defined at $kr_c\phi=-1/\xi$ which is an artifact of the analytic solution failing to be a good approximation. By appropriately choosing $\xi$, we can ensure $-1/\xi >>k\pi r_c$ and hence trust our solution. The radion potential coming from the GW part of the action is then given by
\begin{equation}\label{32}
\begin{split}
V_{GW}(r_c)
=
\int_{0}^{\pi} d\phi\:\frac{1}{r_c}e^{-4kr_c\phi}
\Big(
\partial_\phi\Phi\;\partial_\phi\Phi
+\frac{r_c^2\,\eta}{3}\Phi^3
\Big)
+ e^{-4kr_c\pi} 2\alpha k^{5/2}\Phi(\pi)\;
\end{split}
\end{equation}
By plugging the solution for $\Phi$ (Eq.\ref{31}) into the GW action and integrating over the extra dimension, we get the GW contribution to radion potential. \footnote{Note $\varphi = Ae^{-k\pi r_c}$ is the canonical radion field.} \cite{Chacko1}
\begin{equation}\label{33}
V_{GW}(\varphi)=k^4\left(\frac\varphi A\right)^4
\left[
-\frac{\alpha^2}{4}+\frac{1}{1-\xi\log(\varphi/A)}\left( 2\alpha v +\frac{\alpha^2 \xi}{8}\right) \right] \; 
\end{equation}
The potential is minimized at $\left<\varphi\right>=f$ as
\begin{equation}\label{34}
\begin{split}
-\frac{\alpha^2}{4}+\frac{w}{1-\xi\log(f/A)}=0 \;
\implies \left[\frac{f}{A}\right]^\xi\equiv e^{-\xi kr_c\pi}
\\=e^{1-4w/\alpha^2}+\mathcal{O}(\xi) \; 
\end{split}
\end{equation}
where $w = 2\alpha v +\frac{\alpha^2 \xi}{8}$.\footnote{The linear b.c. parameter $\alpha$ may be tuned accordingly to set the minimum of the GW radion potential at zero.} In terms of the stabilized modulus, this translates to
\begin{equation}\label{35}
k \pi r_c  \approx -\frac{1}{\xi} + \frac{4w}{\alpha^2 \xi} 
           \approx -\frac{1}{\xi} \;
\end{equation} 
To obtain a large warp factor $k\pi r_c$ must be larger than one. This condition is satisfied if $v$ is of order $10^{-2}$ which validates the last approximation in Eq.(\ref{35}). Thus, we see that the stabilized modulus depends directly on $\xi$ (thus $\eta$) and hence its signature. For a positive $\eta$, we get $r_c$ to be negative - hence, unphysical and no modulus stabilization (assuming $v$ to be positive).

% \noindent We notice from Eq.(\ref{33}) that if $\tau$ was neglected, the radion potential would be of the form $\frac{(2\alpha v + \frac{\alpha^2 \xi}{8})}{1+\xi k\pi r_c} k^4 e^{-4k\pi r_c}$. This, as we can see, goes to zero (its minimum) only when $r_c \to \infty$, unlike the previous case where we had a finite $r_c$. This would imply no modulus stabilization. This is especially interesting as it tells us something poignant about the underlying physics - the GW contribution alone cannot stabilize the modulus in the case of a dominating bulk cubic interaction in the absence of gravitational contribution (as was in the original GW prescription). It is indeed the gravitational part of the radion potential which comes to the rescue.
% However, it should be noted that there could be different factors potentially influencing our result - (a) several approximations were made to get to the form for $r_c$ starting from the solution for the scalar field profile and (b) the fact that we ignore the mass term in favour of the cubic interaction while solving is not very sound from a perturbative viewpoint.
\subsection{Bulk quartic interaction}\label{sec4.3}
\noindent We consider the bulk potential where the quartic term dominates,
\begin{equation}\label{36}
    V_b(\Phi) = \frac{1}{4!}\zeta\Phi^4
\end{equation}
We follow the CMS method to obtain the following approximate solution for $\Phi$,
\begin{equation}\label{37}
    \Phi_{approx}(\phi) = -\frac{k^{3/2}\alpha}{4}e^{4kr_c(\phi-\pi)} + \frac{k^{3/2}v}{\sqrt{1+\eta kr_c \phi}}
\end{equation}
where $\eta = \frac{kv^2\zeta}{12}$. The ill-definition of the solution as the denominator vanishes can be tackled in the same fashion as before. Now, plugging this in the GW action and dimensional reduction gives us the radion potential:
\begin{equation}\label{38}
\begin{split}
V_{GW}(r_c)
=
\int_{0}^{\pi} d\phi\:\frac{1}{r_c}e^{-4kr_c\phi}
\Big(
\partial_\phi\Phi\;\partial_\phi\Phi
+\frac{r_c^2\,\zeta}{12}\Phi^4
\Big)
+ e^{-4kr_c\pi} 2\alpha k^{5/2}\Phi(\pi)\;
\end{split}
\end{equation}
Calculating all the terms explicitly and ignoring higher powers of $e^{-4kr_c\pi}$, we get,
\begin{equation}\label{39}
\begin{split}
    V_{GW}(R) = [\int_{0}^{R} dy e^{-4ky} (1+\eta ky)^{-3}
    [\frac{k^5\eta^2 v^2}{4}+\eta k^5 v^2\\ (1+\eta ky)]]
    + e^{-4kR}[-\frac{\alpha^2 k^4}{2} + 2\alpha k^4 v (1+\eta kR)^{-1/2}]
\end{split}
\end{equation}
where we define $R = \pi r_c$ and $y=\phi r_c$. Instead of solving for this humongous expression, we just set its first derivative to zero (we employ Leibniz's rule to the integral part of Eq.\ref{39}) to solve for the minimum.
\begin{equation}\label{40}
    \partial_R V_{GW}(R) = 0
\end{equation}
\begin{equation}\label{41}
\begin{split}
    \implies e^{-4kR} [\frac{k^5\eta^2 v^2}{4}(1+\eta kR)^{-3}+\eta k^5 v^2 (1+\eta kR)^{-2}
    - 8\alpha k^5 v (1+\eta kR)^{-1/2}\\
    - \alpha \eta v k^5 (1+\eta kR)^{-3/2} + 2\alpha^2 k^5] = 0
\end{split}
\end{equation}
As $v$ is a small parameter, $\eta kR << 1$. Thus, we can use $(1+\eta kR)^n \approx 1+n\eta kR$ which gives us,
\begin{equation}\label{42}
\begin{split}
    k\pi r_c [-\frac{3}{4}\eta^3  v^2 - 2 \eta^2  v^2 + 4 \alpha  v \eta + \frac{3}{2}\alpha \eta^2 v ] = \alpha \eta v  + 8 \alpha  v
    - \frac{\eta^2 v^2}{4} - \eta v^2 - 2\alpha^2
\end{split}
\end{equation}
Assuming $\alpha \approx v$, $|\zeta|$ to be small and ignoring terms of order higher than $\mathcal{O}(v^4)$, we get the expression for the stabilized modulus as
\begin{equation}\label{43}
    k\pi r_c  \approx \frac{3}{2\eta} = \frac{18}{kv^2\zeta}
\end{equation}
Thus, we arrive at an expression for the stabilized modulus for a bulk quartic scalar potential. It is indeed easy to see from Eq.(\ref{43}) that if we were to work with an inverted quartic well ($\zeta \rightarrow -\zeta$), $r_c$ acquires a negative value which is unphysical. Hence, just like the standard GW case, we do not get modulus stabilization for the inverted case.\\

\noindent It is to be noted that in our analysis, we tuned $|\zeta|$ to be small (upto cubic coefficients were even smaller).  This indicates a light radion. However, coefficients of all the terms in the bulk potential could be small as well which would physically correspond to the GW scalar field being a pseudo-Goldstone boson.
\subsection{Bulk double-well potential}\label{sec4.4}
\noindent Let the potential for the GW scalar in the bulk have the form
\begin{equation}\label{44}
    V_b(\Phi) = \lambda(\Phi^2 - b^2)^2
\end{equation}

\begin{figure}[h]\label{f1}
    \centering
    \includegraphics[width=0.8\linewidth]{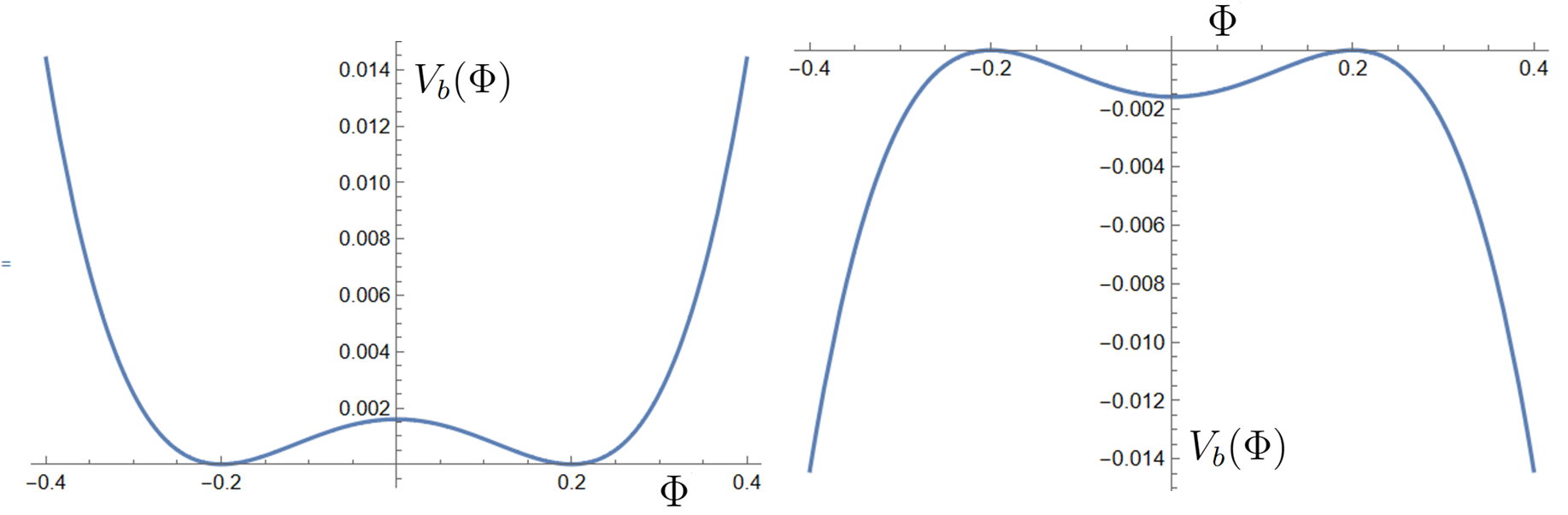}
    \caption{Form of the bulk potential $V_b(\Phi)$ with $b=0.2$ for $\lambda = 1$ (left) and $\lambda=-1$ (right)}
\end{figure}

\noindent We again follow the CMS method for this more general form of bulk potential as the classical equation of motion for $\Phi$ is certainly difficult to solve analytically.
\begin{equation}\label{45}
    \partial^2_\phi \Phi - 4kr_c\partial_\phi \Phi + 4\lambda r_c^2 \Phi (\Phi^2 - b^2) = 0
\end{equation}
\begin{figure*}\label{f2}
    \centering
    \includegraphics[width=0.9\linewidth]{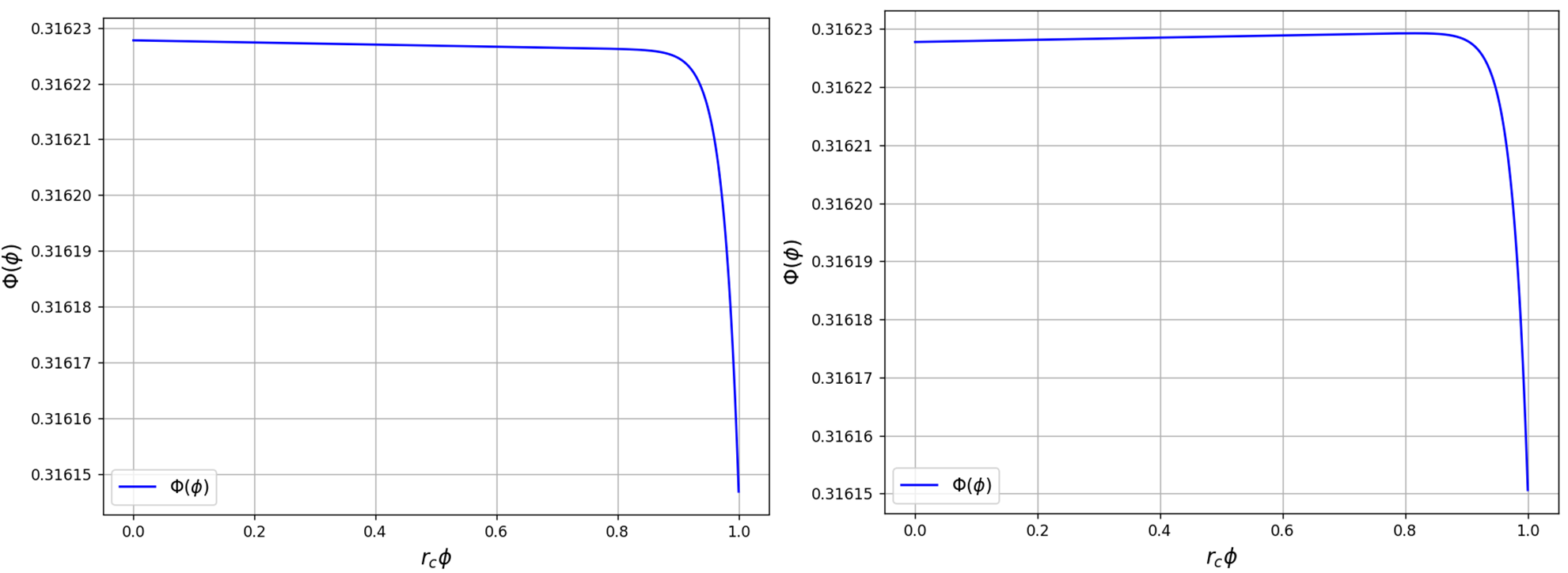}
    \caption{Scalar field profile for parameter values $b=0.2, v=0.01, k=10, \alpha = 10^{-5}$ with $\lambda = 0.001$ (left) and $\lambda  = -0.001$ (right)}
\end{figure*}\\
We use Eq.(\ref{23}) to find the OR solution for $\Phi$ which takes the form
\begin{equation}\label{46}
    \Phi_{OR}(\phi) = \frac{b}{\sqrt{1-e^{2b^2(a\phi+d)}}}
\end{equation}
where $a=-\frac{r_c\lambda}{k}$ and $d=\frac{1}{2b^2}\ln[1-\frac{b^2}{k^3v^2}]$. The integral constant \enquote*{d} is found out by using the boundary condition
\begin{figure*}\label{f3}
    \centering
    \includegraphics[width=1.0\linewidth]{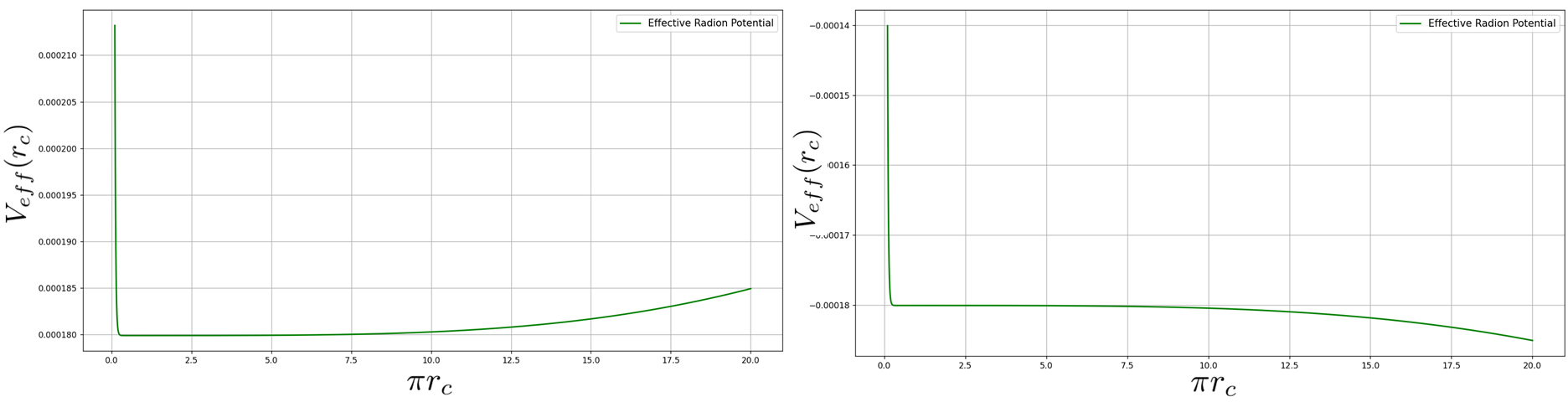}
    \caption{Effective radion potential (double well) for parameter values $b=0.2, v=0.01, k=10, \alpha = 10^{-5}$ with $\lambda = 0.001$ (left) and $\lambda  = -0.001$ (right)}
\end{figure*}
$\Phi(\phi=0) = k^{3/2}v$. Matching with the BR solution Eq.(\ref{24}), we get the approximate solution for $\Phi$ as
\begin{equation}\label{47}
    \Phi_{approx}(\phi) = -\frac{k^{3/2}\alpha}{4}e^{4kr_c(\phi-\pi)} + \frac{b}{\sqrt{1-e^{2b^2(a\phi+d)}}}
\end{equation}

\noindent Note that this solution is ill-defined when $a\phi+d=0$ which implies $r_c\phi=\frac{k}{2b^2\lambda}\ln[1-\frac{b^2}{k^3v^2}]$. Now, $(1-\frac{b^3}{k^3v^2})$ is certainly less than 1, k being positive, resulting in a negative value of $r_c$ which is unphysical. Thus, the solution is well defined along the entire positive $r_c$ (physical) axis. As with the original GW scheme, we fine-tune the brane tensions to make the gravitational contribution to the radion potential vanish. So, the only contribution is of the GW type. Hence, we take the scalar field profile given by Eq.(\ref{47}) and plug it in Eq.(\ref{19}) with the brane potentials:
\begin{eqnarray}
    V_h &=& \lambda_h (\Phi^2 - k^3 v^2)^2 \label{48}\\
    V_v &=& 2 \alpha k^{5/2} \Phi \label{49}
\end{eqnarray}
We then integrate over the extra dimension i.e. $\phi$ to get the expression for the effective radion potential. It is quite cumbersome to get the analytic expression. So, we resort to the numerical results (see Fig. \hyperref[f3]{3}) which do provide the necessary insights.\\

\noindent What we get is that the effective potential has a global minimum for a double well bulk potential i.e. the modulus gets stabilized to a non-trivial non-zero value, which for our benchmark choice of parameters in Fig. \hyperref[f3]{3} turns out to be 0.31. As a result, $k\pi r_c \approx31$ which very well solves the hierarchy problem. However, for the inverted double well, the modulus gets destabilized (no minimum).

\subsection{Bazeia-Furtado-Gomes type bulk potential}\label{sec4.5}
\noindent Bazeia, Furtado and Gomes (BFG); in Ref.\cite{DBazeia_2004}, introduced a set of bulk scalar potentials which are generated from a class of superpotentials of the form
\begin{equation}\label{50}
    W_p(\Phi)=\frac{2p}{2p-1}\,\Phi^{\frac{2p-1}{p}}-\frac{2p}{2p+1}\,
\Phi^{\frac{2p+1}{p}}
\end{equation}
where p is an odd integer, and the bulk potential given by [taking $\kappa^2 = 2$]:
\begin{equation}\label{51}
    V_{b,p}(\Phi)=\frac18 \left(\frac{dW_p}{d\Phi}\right)^2-\frac13 W^2_p
\end{equation}

\noindent The metric ansatz we use is
\begin{equation}\label{52}
    ds^2 = e^{2A(y)}\eta_{\mu\nu}dx^{\mu}dx^{\nu}-dy^2
\end{equation}
\begin{figure*}\label{f4}
    \centering
    \includegraphics[width=0.7\linewidth]{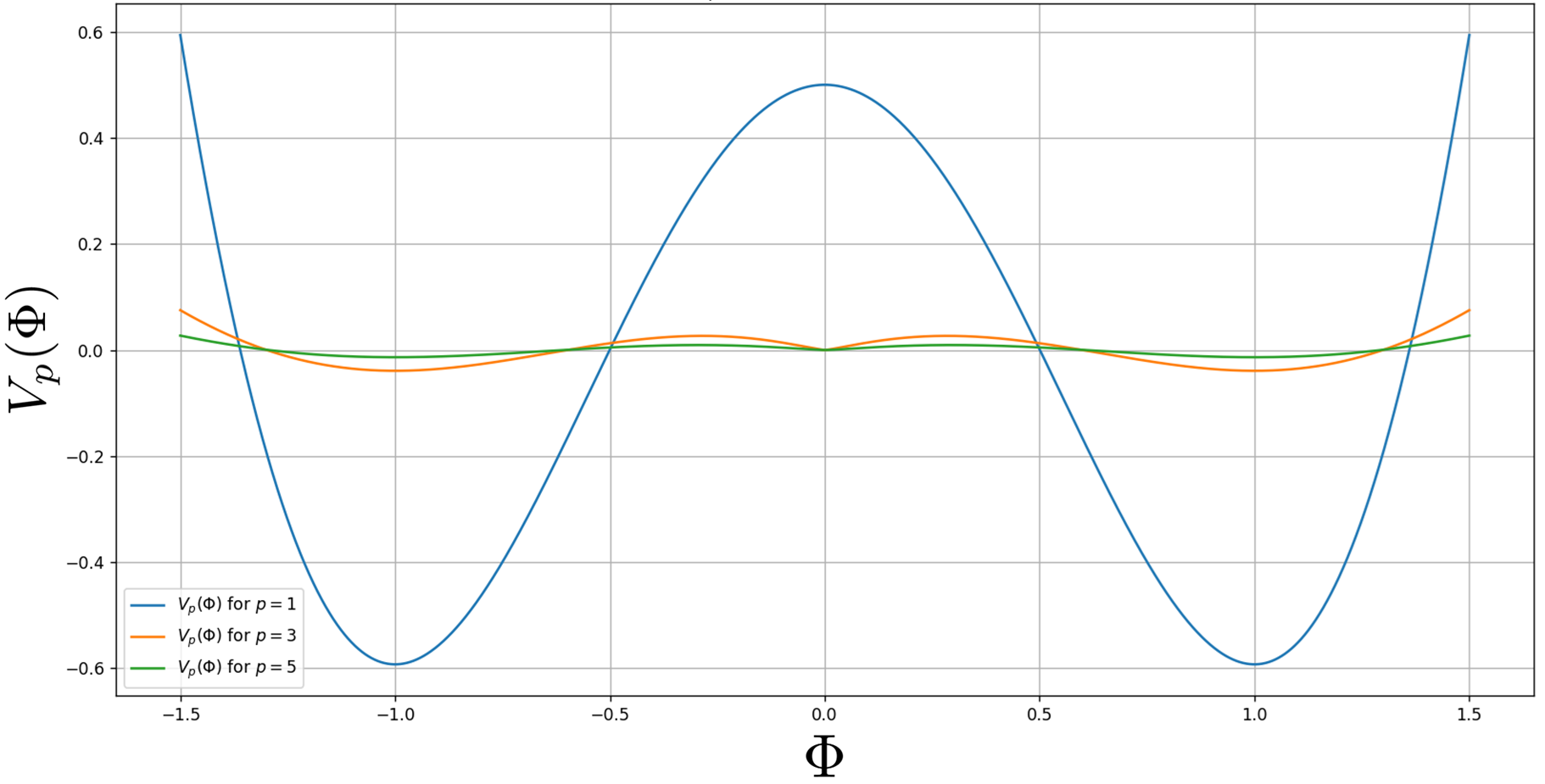}
    \caption{BFG potentials for different p values}
\end{figure*}

\noindent The speciality of these potentials is that the coupled gravity-scalar system gives rise to thick brane solutions. What we do here is take such forms of bulk potentials and plug them in the RS-I background with proper brane potentials - hence solving for the scalar profile and backreacted metric using the superpotential approach of Ref. \cite{Freedman} (see Appendix \ref{appa}). The next step is to calculate the effective radion potential and check the behaviour for the entire class of superpotentials. A final question is if we can establish a relation between the value of the stabilized modulus ($r_c$) [if it gets stabilized] and the parameter p, which is addressed in Appendix \ref{appb}.\\

\noindent Following the steps as in Appendix \ref{appa}, we obtain the following solutions for the scalar profile and the warp factor \cite{DBazeia_2004}:

\begin{equation*}
    \Phi_p(y)=\tanh^p\left(\frac{y}{p}\right)
\end{equation*}
\begin{equation}\label{53}
\begin{split}
    \noindent A_p(y)\!\!=\!\!-\frac13\frac{p}{2p+1}\tanh^{2p}\left(\frac{y}{p}\right)-
\frac23\left(\frac{p^2}{2p-1}-\frac{p^2}{2p+1}\right)
\biggl{\{}\ln\biggl[\cosh\left(\frac{y}{p}\right)\biggr]-
\sum_{n=1}^{p-1}\frac1{2n}\tanh^{2n}\left(\frac{y}{p}\right)\biggr{\}}
\end{split}
\end{equation}

\noindent One should notice that, unlike all the cases till now, this is an exact solution involving the backreaction. The additional inputs that we need to give are the brane potentials which are precisely determined by Eqs. (\ref{a10}) and (\ref{a11}). Since the scalar field vanishes at the hidden brane and so does the superpotential evaluated there, $\lambda_h$ does not contribute to the radion potential. On the other hand, the $\lambda_v$ contribution to the radion potential is precisely given by $-W(\Phi_v).e^{{A_{p}(vis)}}$. Hence, putting in the scalar field solution and modified warp factor into the combined gravity-scalar action, integrating over the extra dimension and adding the contribution due to brane potentials; we get the effective radion potential \footnote{The Ricci scalar in the BFG convention is given by $R = 8A"+20A'^2$. This is incorporated in the numerics.}. In this scenario as well, the calculations are quite cumbersome to do analytically, especially when our sole purpose is to determine the presence of modulus stabilization. So, we resort to a numerical approach.\\
\begin{figure*}\label{f5}
    \centering
    \includegraphics[width=0.8\linewidth]{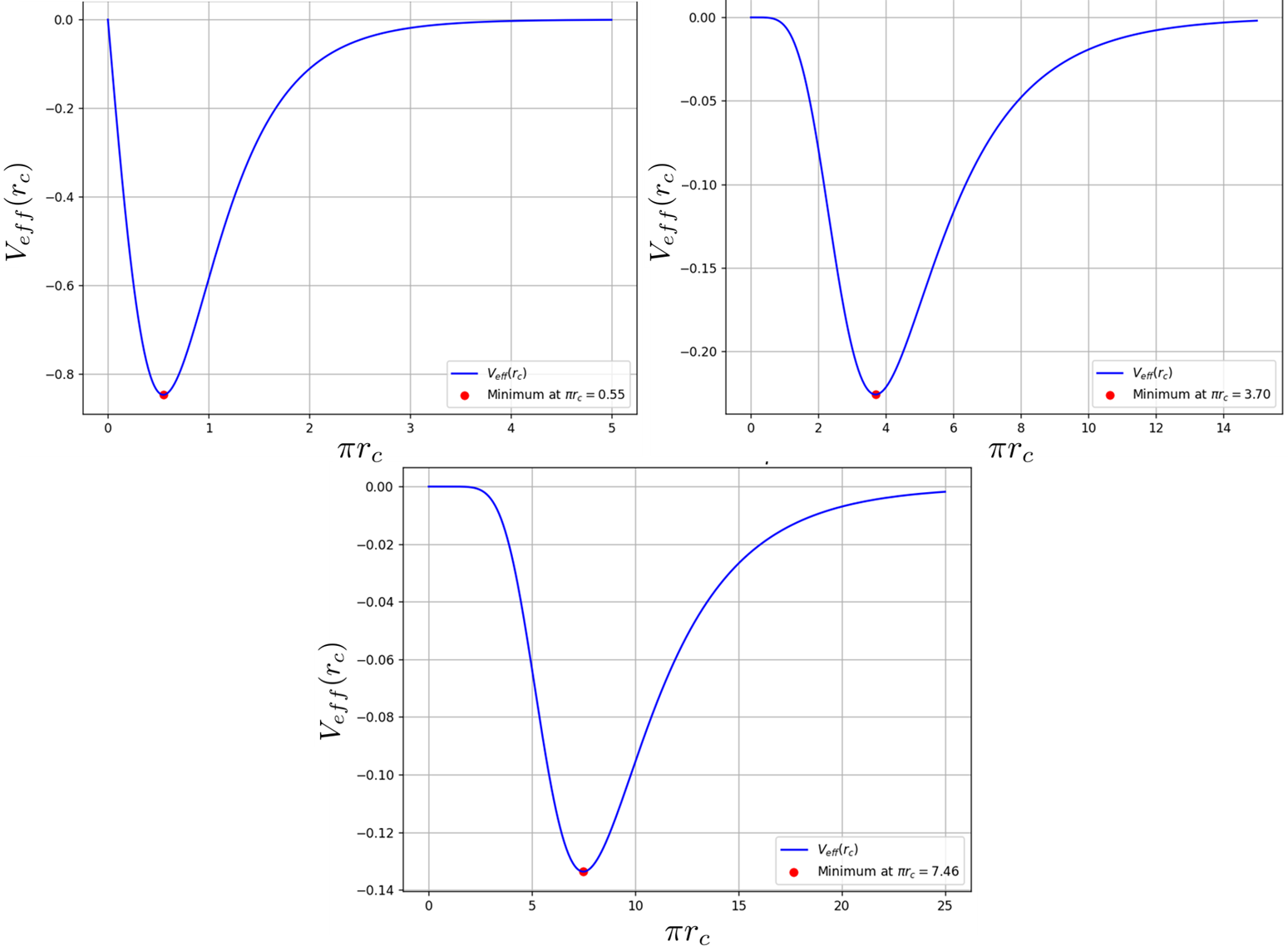}
    \caption{Effective radion potential for p = 1 (top left), p = 3 (top right) and p = 5 (bottom)}
\end{figure*}

\noindent From Fig. \hyperref[f5]{5}, we see that one gets modulus stabilization for BFG type bulk scalar potentials at least for p till 19 which has been checked numerically - The computational resources required for the numerical calculations increase as the value of p increases. But, since the potentials for all further p values retain the same structure (see Fig. \hyperref[f5]{5}), we conjecture that one gets modulus stabilization for the entire class of BFG type bulk scalar potentials.\\

% \noindent As the scalar field solution is a tan hyperbolic raised to an odd power, it has values between plus and minus one. This restricts the domain of the bulk potentials and hence, the bulk potentials are bounded from below by their values at $\pm 1$. Given the nature of our argument for all the cases till now, we would have naturally expected modulus stabilization in these scenarios which is indeed what we get.

\subsection{Mishra-Randall ansatz}\label{sec4.6}
\noindent Recently, Mishra and Randall \cite{Randall_2023} explored the effect of including bulk interaction terms for the GW scalar on RS cosmology. In particular, they were interested in the form of the beta function ($\beta = 1/T$), where T is the
\begin{figure*}\label{f6}
    \centering
    \includegraphics[width=0.7\linewidth]{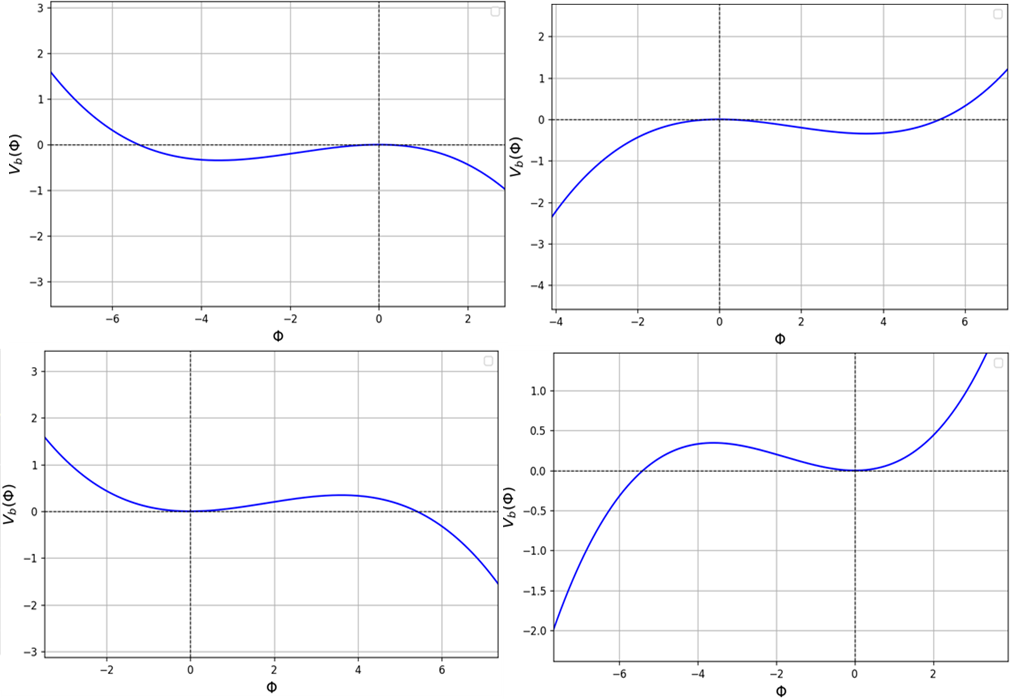}
    \caption{GW bulk potentials [($\epsilon_2<0,\epsilon_3<0$);($\epsilon_2<0, \epsilon_3>0$); ($\epsilon_2>0, \epsilon_3<0$); ($\epsilon_2>0,\epsilon_3>0$)]: from left to right}
\end{figure*}
\begin{figure*}\label{f7}
    \centering
    \includegraphics[width=0.7\linewidth]{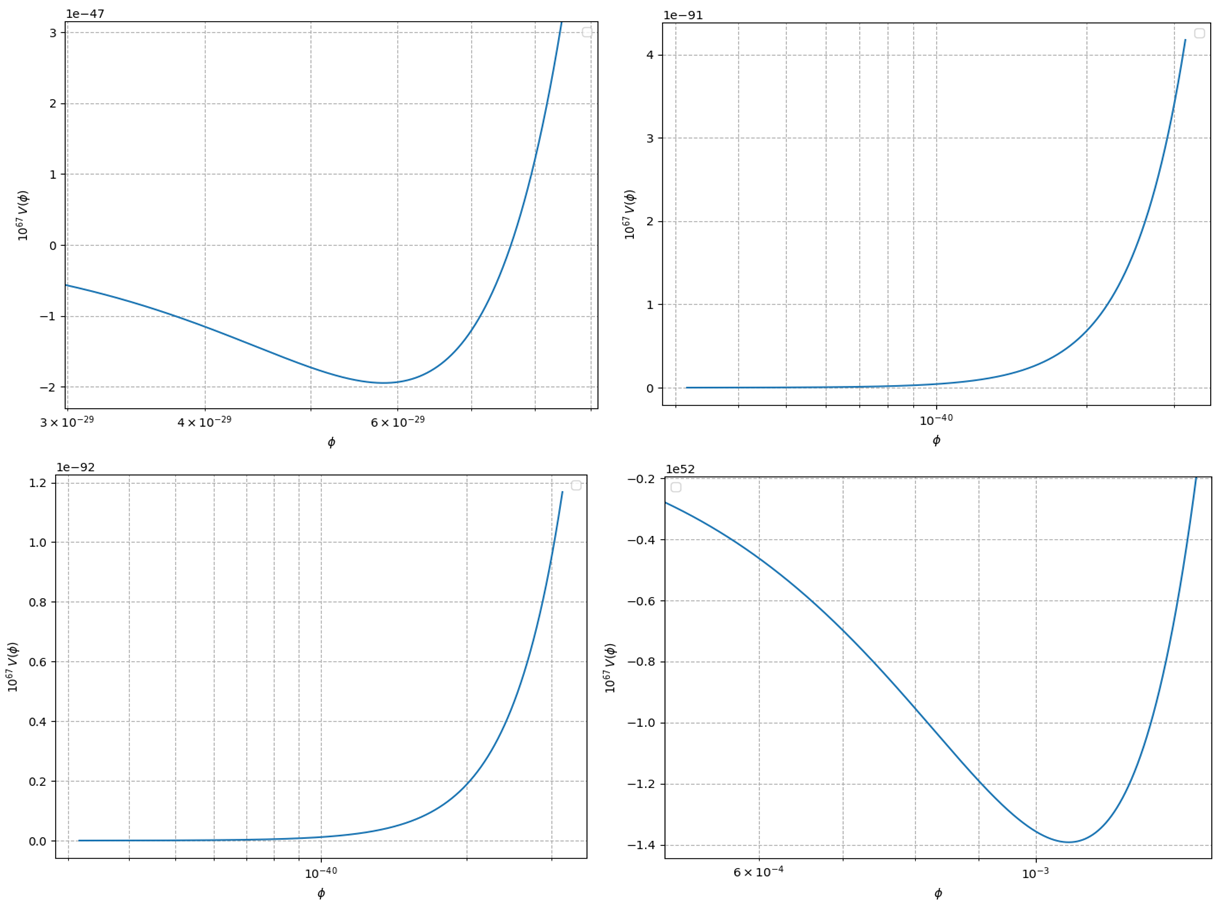}
    \caption{Radion potentials [($\epsilon_2<0,\epsilon_3<0$);($\epsilon_2<0, \epsilon_3>0$); ($\epsilon_2>0, \epsilon_3<0$); ($\epsilon_2>0,\epsilon_3>0$)]: from left to right. $10^{67} V(\varphi)$ labels the vertical axis and $\varphi$ labels the horizontal axis.}
\end{figure*}
% \begin{figure*}\label{f8}
%     \centering
%     \includegraphics[width=1.0\linewidth]{mishrandall_wec.png}
%     \caption{WEC check [($\epsilon_2<0,\epsilon_3<0$);($\epsilon_2<0, \epsilon_3>0$); ($\epsilon_3<0,\epsilon_2>0$); ($\epsilon_2>0,\epsilon_3>0$)]: from left to right}
% \end{figure*}
black hole temperature whose origin lies in its Hawking radiation. In the RS background, this function could take up any value. For no bulk interaction of the GW scalar, the beta function is a constant, and higher-order terms in the GW bulk potential correspond to higher-order terms in the beta function. With this in mind (to model the additional terms in beta), they considered an ansatz of the form:
\begin{equation}\label{54}
    V_b(\Phi) = 
    2\epsilon_2\Phi^2+\frac43\epsilon_3\Phi^3
\end{equation}
The brane potentials that they consider are of the form
\begin{equation}\label{55}
    \lambda_{hid}(\Phi) = \beta_{hid}(\Phi-v_{hid})^2\:,\;\:\beta_{hid}\to\infty\:,
     \\
    \:\lambda_{vis}(\Phi) = 2\alpha_{vis}\Phi
\end{equation}
The scalar field profile, as given by the CMS scheme is
\begin{equation}\label{56}
    \Phi_\text{RS}(r) = -\frac{\alpha_{vis}}{4} e^{4(r-r_{vis})} + \frac{v_{hid} e^{-\epsilon_2 r}}{1 + v_{hid} \epsilon_3 \left(\frac{1-e^{-\epsilon_2 r}}{\epsilon_2}\right)}
\end{equation}
with $0\leq r \leq r_{vis}$. Plugging everything into the GW action and integrating over the fifth dimension gives the following form for the radion potential \cite{Randall_2023}
\begin{equation}\label{57}
\begin{split}
    V_{GW}(\varphi) = \varphi^4(A+ B\frac{\lambda\varphi^{\epsilon_2}}{1-\lambda\varphi^{\epsilon_2}} + C(\frac{\lambda\varphi^{\epsilon_2}}{1-\lambda\varphi^{\epsilon_2}})^3 - D\log (1-\lambda\varphi^{\epsilon_2}))
\end{split}
\end{equation}
with
\begin{equation}\label{58}
\begin{split}
    \qquad \varphi = e^{-r_{vis}},\: \lambda = \frac{v_\text{hid} \epsilon_{32}}{1+ v_\text{hid}\epsilon_{32}},\:
    \epsilon_{32} = \frac{\epsilon_3}{\epsilon_2},
    \: A= \frac{1}{768}\epsilon_3 \alpha_{vis}^2 - \frac{\lambda}{1-\lambda} + \log(1-\lambda),\\\: B = -\frac{1}{32}\epsilon_2\alpha_{vis}^2-\frac{1}{2}\frac{\epsilon_2}{\epsilon_3}\alpha_{vis},\: C = \frac{1}{6}\frac{\epsilon_2^3}{\epsilon_3^2},\: D = \frac12\frac{\epsilon_2}{\epsilon_3}\alpha_{vis}
\end{split}
\end{equation}
We work with a specific choice of parameters, which was part of the benchmark choice in Ref.\cite{Randall_2023} as well: $\epsilon_2 = -1/25, \epsilon_3 = -1/90, \alpha_{ir} = 5/2, v_{uv} = 1/5$. The resulting radion potential does have a non-trivial minimum. We now work with the same magnitudes for $\epsilon_2$ and $\epsilon_3$ but with different sign combinations: namely one of $\epsilon_2$ or $\epsilon_3$ being positive or both being positive. In the first case, we get a monotonically increasing radion potential with a minimum at $\varphi = 0$, corresponding to $r_{vis} \to \infty$ but in the second case, the radion potential has a non-trivial minimum. Thus, we see that the essential condition for radion stabilization in this scenario is that the quadratic and cubic coefficients in the bulk scalar potential must share the same signature, or simply, $\epsilon_{32}>0$.
% We also checked the weak energy condition (WEC) in each of the cases to see if there are some implications - however, the results do not seem to be hinting at some underlying physics [Fig.8].\\

% \noindent We get modulus stabilization for the $\epsilon_3=0, \epsilon_2<0$ case as is shown in Ref.\cite{Randall_2023}. However, the interesting case is $\epsilon_2\to0$ i.e. only cubic contribution. We see from Eq.(\ref{58}) that in this limit, $a_2=2\alpha_{vis},a_3=0, \epsilon_{32}\to \infty$ and hence, $\lambda \to 1$. Plugging these in Eq.(\ref{57}), we see that the third term becomes zero while the second term is non-zero. The most important thing to notice here is that as all the $\phi$ contributions within the brackets were raised to the power $\epsilon_2$, we are left with no $\phi$ dependance for the quartic coefficient now - it is a constant. Thus, the radion VEV is either zero or driven to infinity depending on the sign of this constant implying no modulus stabilization irrespective of the sign of $\epsilon_3$. This is certainly what we expected as this was our conclusion when we independently examined the cubic bulk potential GW scenario without the gravitational contributions. Thus, it serves as a good check that our results match under proper limits.
\subsection{General comments}\label{sec4.7}
Let us summarize all the cases that we discussed till now. (a) \textit{No bulk interaction}: only a positive mass-squared bulk scalar ensures radion stabilization. (b) \textit{Bulk cubic interaction}: only a negative cubic coefficient in the bulk potential stabilizes the radion. (c) \textit{Bulk quartic interaction}: only a positive quartic coefficient in the bulk potential results in modulus stabilization. (d) \textit{Bulk double-well interaction}: a bulk double-well potential stabilizes the radion but an inverted well does not. (e) \textit{BFG type bulk potential}: ensures radion stabilization for any BFG superpotential. (f) \textit{Mishra-Randall (MR) ansatz}: modulus stabilization happens only when quadratic and cubic coefficients share the same signature. Except for Sec.\ref{sec4.1} and the BFG case, all the cases that we discussed had a GW-type Dirichlet boundary condition on the hidden brane and a linear brane potential on the visible brane, ensuring a consistent comparison. The equivalent analysis for Sec.\ref{sec4.1} with CMS boundary conditions can be found in Sec.\ref{sec5.2}. The BFG case was carried out only to demonstrate a non-trivial backreacted geometry. It is known that in the superpotential method, one usually fixes the values of the bulk scalar on the Planck and TeV brane, and this stabilizes the radion - thus it might appear that there is no point in studying such a scenario. But our situation was a bit modified - the scalar field value on the Planck brane was fixed (which serves as an initial condition for the scalar solution), but on the visible brane was not. Instead upon plugging in the scalar profile in the 5D action, we performed dimensional reduction as previously to get the radion potential. The fact that we still got radion stabilization is thus no longer trivial.\\  

\noindent The question now is - what possible conjectures could one draw from these cases? For bulk potentials with a local (and no global) minimum, the radion may or may not be stabilized as per the inverted double-well case and the different scenarios of MR ansatz. The same goes for bulk potentials unbounded above and below - both for the cubic case and the variants of MR ansatz, selective modulus stabilization was observed. Thus, though it might be worthwhile to explore general restrictions on the parameters under these conditions (case-specific restrictions were already discussed), there is nothing one could conjecture purely from the qualitative nature of the bulk potential.\\

\noindent A consistent observation is that anywhere the bulk potential had a global minimum, it resulted in radion stabilization, hinting at a generalization of this statement. This was seen to be true for the BFG potentials as well since they have a global minimum, though under different boundary conditions. Again, all the cases where the bulk potential had a global maximum but no global minimum failed to stabilize the radion - can this inference be generalized? A subset of these cases is infact those unstable bulk scalars which violate the BF bound - is there a further connection to scalar field stability? These are the conjectures that rely entirely on the qualitative nature of the bulk scalar potential and are addressed in the next section. 

\section{Addressing the correspondence: consequences of an extremum in the bulk potential}\label{sec5}
% \noindent Here, we discuss two classes of the GW bulk potential (the potential satisfying certain properties) that clearly indicate the one-one correspondence between the nature of the extremum in the bulk potential and successful modulus stabilization. It is worthwhile to note that the below proofs include some approximations and hence, we would be critical and say that the correspondence is not exact - rather the possibilty of it being valid is incredibly high.
\subsection{Global minimum for $V_b(\Phi)$}\label{sec5.1}
\noindent The existence of a global minimum for the bulk scalar potential has strong implications for the OR solution of $\Phi$ (see Eq.\ref{23}). Particularly, the OR equation of Eq.(\ref{23}) indicates the existence of a stable fixed point for $\Phi_{OR}$ at $\Phi_0$ - the scalar field value at which $V_b(\Phi)$ attains its global minimum. As a result, the profile of $\Phi_{OR}$ is such that it gets a lot of contribution from the global minimum well, around $\Phi_0$. This allows us to treat the bulk potential as an expansion around its minimum. As we are interested in the nature of the potential, we can take the global minimum at the origin i.e. $\Phi_0=0$ and $V(\Phi_0)=0$. The rescaling essentially implies corrections to the bulk cosmological constant which is assumed to still remain negative ($AdS_5$). We alternatively call $V_b(\Phi)$ as $V(\Phi)$ and $V_b''(\Phi)|_{\Phi=\Phi_0}$ as $V''$ for brevity.
\begin{equation}\label{59}
    V(\Phi) = V(\Phi_0) + \frac{1}{2}V''(\Phi)|_{\Phi=\Phi_0}(\Phi-\Phi_0)^2 = \frac{1}{2}V''(\Phi)|_{\Phi=\Phi_0}\Phi^2
\end{equation}
neglecting the higher order terms. Note that $V''\geq0$ as it is a minimum but for this proof, we restrict to $V''>0$. In the discussion below, by \enquote{conditions}, we mean boundary conditions and brane couplings.\\

\noindent We note that the GW scalar field $\Phi$ in five dimensions has mass dimension $[\Phi] = 3/2$. As a result, higher-order terms in the bulk potential, such as $\Phi^n$ for $n\geq 4$ are suppressed by powers of the cutoff scale - the 5D Planck scale $M$. For instance, the operator $\Phi^4$ has mass dimension 6 and thus appears with a coefficient of order $1/M$, while higher order terms are even more suppressed. Therefore, in the effective field theory regime where $\Phi<<M^{3/2}$, the potential can reliably be truncated at quadratic (or at most cubic) order.\\

\noindent With GW conditions [as defined in Sec.\ref{sec2.3}], the analysis is exactly the same as Sec.\ref{sec2.3} except $V''(\Phi)|_{\Phi=\Phi_0}$ representing the squared mass of the bulk scalar. Hence, we can say- \textit{under GW conditions, if the bulk scalar potential has a global minimum satisfying $V''>0$, then it implies modulus stabilization.} However, an extensive calculation with the CMS conditions is yet to be worked out which is what we do now. It must be noted that we go for a full expansion of the radion potential unlike Ref.\cite{Chacko1} which calculates only to the leading order in $v$ albeit incorporating gravitational sector contribution.\\

\noindent We write the OR solution as
\begin{equation}\label{60}
    \Phi_{OR} = \Phi_0 + \epsilon(\phi) = \epsilon(\phi)
\end{equation}
Plugging these expressions into the OR equation $\frac{d\Phi_{OR}}{d\phi} = -\frac{r_c}{4k}V'(\Phi)$, we get the following solution for $\Phi_{OR}$,
\begin{equation}\label{61}
    \Phi_{OR}(\phi) = \epsilon_0 e^{-r_cV''\phi/4k}
\end{equation}
with $\epsilon_0$ being the integration constant which is determined by the boundary condition $\Phi(0) = k^{3/2}v$. Finally, combining with the BR solution, the resulting scalar field profile is given by \footnote{Parameters $\alpha$ and $v$ are taken to be positive.}
\begin{equation}\label{62}
    \Phi(\phi) = -\frac{k^{3/2}\alpha}{4}e^{4kr_c(\phi - \pi)} + k^{3/2}ve^{-r_cV''\phi/4k}
\end{equation}
The GW part of the radion potential is obtained by using this solution in Eq.(\ref{19}) with the choice for visible brane potential determined by Eq.(\ref{20}).
\begin{equation}\label{63}
\begin{split}
    V_{GW}(r_c) = \frac{k^4\alpha^2}{4}e^{-8kr_c\pi}(e^{4kr_c\pi}-1)-\frac{(V'')^2}{8k}\frac{k^3 v^2}{V''+8k^2}(e^{-(\frac{r_cV''}{2k} + 4kr_c)\pi}-1)\\-k^{4}\alpha v e^{-4kr_c\pi}(e^{-r_cV''\pi/4k}-1)-2k^4v^2\frac{V''}{V''+8k^2}(e^{-(\frac{r_cV''}{2k} + 4kr_c)\pi}-1)\\-\frac{k^4\alpha^2}{2}e^{-4kr_c\pi}+2k^4\alpha v e^{-(\frac{r_cV''}{4k} + 4kr_c)\pi}
\end{split}
\end{equation}
We now calculate the derivatives of the radion potential,
\begin{equation}\label{64}
\begin{split}
    V'_{GW}(r_c) = k^5 \alpha^2\pi e^{-4kr_c\pi}-4k^5\alpha \pi v e^{-4kr_c\pi}+2k^5\alpha^2\pi e^{-8kr_c\pi}\\+\frac{k\pi}{16}v^2(V'')^2 e^{-(\frac{r_cV''}{2k}+4kr_c)\pi}+k^4\alpha\pi v(4k+\frac{V''}{4k})e^{-(\frac{r_cV''}{4k}+4kr_c)\pi} \\+k^3v^2 \pi V'' e^{-(\frac{r_cV''}{2k}+4kr_c)\pi}-2\alpha k^4 v \pi (4k+\frac{V''}{4k})e^{-(\frac{r_cV''}{4k}+4kr_c)\pi}
\end{split}
\end{equation}
The fact that $4kr_c>>1$ allows us to neglect the $e^{-8kr_c\pi}$ term. If $V'' \gtrsim 16k^2$, then one obtains $e^{-(\frac{r_cV''}{4k}+4kr_c)\pi} \lesssim e^{-8kr_c\pi}$ and $e^{-(\frac{r_cV''}{2k}+4kr_c)\pi} \lesssim e^{-12kr_c\pi}$. Consequently, it does not make sense, in general, to neglect the $e^{-8kr_c\pi}$ term alone, while simultaneously retaining the other terms. Thus, neglecting the $e^{-8kr_c\pi}$ term with respect to all the other terms in Eq. 5.6 is justified only for $V''<16k^2$. This upper bound for $V''$ has its justifications - firstly, $V''$ corresponds to mass squared of the bulk scalar and we take $\frac{V''}{k^2}<<1$, where k is the AdS curvature scale to ensure that backreaction can be neglected and we can solve for the bulk scalar in the original RS background (which is what we do here). Secondly, in the context of the AdS-Swampland conjecture \cite{lust2019ads}, a consistent theory of quantum gravity is expected to forbid parametrically light or isolated AdS vacua. Specifically, the AdS curvature scale $k$ (with $\Lambda_b \sim -k^2$) is conjectured to be tied to the mass scale of an infinite tower of states, $m_{\text{tower}} \sim |\Lambda_b|^{1/2} \sim k$. Consequently, the effective field theory description breaks down when scalar masses in the bulk significantly exceed the AdS curvature scale, implying a soft upper bound $V'' \lesssim \mathcal{O}(1)\,k^2$.\\

\noindent At the extremum (supposing it exists) say $r_c^*$, Eq.(\ref{64}) vanishes. The expression for $r_c^*$ is given by,
\begin{equation*}
\begin{split}
   e^{-\frac{r_c^*V''\pi}{4k}} = \frac{k^3\alpha v (4k +\frac{V''}{4k}) \pm \sqrt{X}}{V''(2kv^2+\frac{v^2V''}{8})}
\end{split}
\end{equation*}
where
\begin{equation*}
\begin{split}
   X = (V'')^2(k^4\alpha v^3 - \frac{3}{16}k^4\alpha^2 v^2) + V''(16k^6\alpha v^3 - 2k^6\alpha^2 v^2) + 16k^8 \alpha^2 v^2
\end{split}
\end{equation*}
The constraint on the $\alpha, v, V''$ parameter space comes from two requirements: (a) $X \geq 0$ and (b) $0<e^{-\frac{r_c^*V''\pi}{4k}}<1$ to ensure the modulus value $r_c^*>0$ (as $V''>0$). For $V''<0$ (see Sec.5.2), the second criterion changes as  $1<e^{-\frac{r_c^*V''\pi}{4k}}<\infty$. Thus, we can have either zero, one or two extrema. The second derivative of the radion potential at an extremum $r_c^*$ is then given by
\begin{equation}\label{65}
\begin{split}
    V''_{GW}(r_c)|_{r_c^*}= \frac{k^2\pi^2\alpha v}{16}V''(V''+16k^2)e^{-(\frac{r_c^*V''}{4k}+4kr_c^*)\pi}
\end{split}
\end{equation}
where we have dropped $\mathcal{O}(e^{-(\frac{r_cV''}{2k}+4kr_c)\pi})$ terms in favour of $\mathcal{O}(e^{-(\frac{r_cV''}{4k}+4kr_c)\pi})$ terms and used the fact that Eq.(\ref{64}) vanishes at $r_c^*$. As $V''>0$, we can see that Eq.(\ref{65}) is strictly positive as long as $\alpha v >0$ (we can always impose this by boundary conditions). Thus, the extremum is surely a minimum.\\

\noindent \textit{Under CMS conditions, if the bulk scalar potential has a global minimum satisfying $V''>0$ and the corresponding radion potential has an extremum for a positive value of $r_c$, then the extremum is certainly a minimum of the radion potential, implying modulus stabilization.}

\subsection{Global maximum but no global minimum for $V_b(\Phi)$}\label{sec5.2}
\noindent The analysis follows as same except now, the perturbative solution for the scalar field is a runaway solution as the point of maximum is an unstable fixed point of the system and there exists no global minimum. The bulk potential can be approximated as same except $V''\leq0$ - for the proof, we take it to be $V''<0$.\\

\noindent With GW conditions, the analysis is again as given in Sec.\ref{sec4.1}. We deal with a negative mass term which essentially results in modulus destabilization as we showed previously. To put it in words; \textit{under GW conditions, if the bulk scalar potential has a global maximum satisfying $V''<0$, then the corresponding radion potential has no extremum for a positive value of $r_c$ implying no modulus stabilization.}. We now move onto the CMS conditions.\\

\noindent $V''<0$ allows us to only consider the terms involving $e^{-\frac{r_cV''\pi}{4k}}$ and $e^{-\frac{r_cV''\pi}{2k}}$ in Eq.(\ref{64}) as they are increasing exponentials, $V''$ being negative. Supposing an extremum exists, then the second derivative of the radion potential at the extremum $r_c^*$ is given by
\begin{equation}\label{66}
    V''_{GW}(r_c)|_{r_c^*} = -\frac{k^2\pi^2 \alpha v}{16}V''(V''+16k^2) e^{-(\frac{r_c^*V''}{4k}+4kr_c^*)\pi}
\end{equation}
where we have used the vanishing of the first derivative at extremum. $V''$ being negative and ensuring $\alpha v >0$ as in the previous proof, we have two cases: (a) $V''<-16k^2$ indicates a maximum and (b) $-16k^2<V''<0$ indicates a minimum. Thus we get to the following theorem,\\

\noindent \textit{Under CMS conditions, if the bulk scalar potential has a global maximum but no global minimum satisfying $V''<0$, and the corresponding radion potential has an extremum for a positive value of $r_c$, then the extremum is a maximum of the radion potential if $V''<-16k^2$ or a minimum of the radion potential if $-16k^2<V''<0$ implying modulus destabilization or stabilization respectively.}\\

\noindent A point to note here is that the BF condition $\frac{m^2}{k^2}+4>0$ translates to $\frac{V''}{16k^2}>-\frac{1}{4}$. Thus if the BF condition were to be satisfied i.e. the bulk field were to be stable, then the $\frac{V''}{16k^2}<-1$ is unphysical and hence, the radion potential can only possess a minimum. However, if the bulk scalar is taken to be unstable, then the resulting radion potential can possess either a maximum or a minimum. For the $V''=0$ case, the same analysis follows except then, we would have to take the first non-zero term in the Taylor expansion of the bulk potential.\\

\noindent We can thus conclude that if the bulk scalar potential has a global minimum and the corresponding radion potential possesses an \textit{appropriate} \footnote{\textit{Appropriate} means (here and further) that the modulus value $r_c$ should be positive.} extremum, then radion stabilization is guaranteed.
% Owing to witnesses of radion stabilization where the bulk potential lacked a global minimum, we can then comment that the existence of a global minimum in the bulk potential is sufficient but not a necessary condition for modulus stabilization in cases where the corresponding radion potential possesses an \textit{appropriate} extremum.
However, for bulk scalar potentials with a global maximum and no global minimum, and the corresponding radion potential having an \textit{appropriate} extremum, we saw that modulus stabilization may or may not be achieved depending on the choice of our parameters for the bulk potential. In this context, if stability (satisfying the BF bound) of the bulk field is further taken into account, then stable fields will always stabilize the radion, and unstable fields may or may not stabilize the radion. An important takeaway is thus the fact that failing to stabilize the radion can only result from an unstable bulk field.

% Note that these deductions are consistent with but do not explain the case studies for bulk cubic interaction and the Mishra-Randall ansatz. \\

% \footnote{For the CMS results, we assumed a positive $\alpha v$.}

\section{General mathematical conditions for the bulk potential to ensure modulus stabilization}\label{sec6}
\noindent In this section, we will frequently use tools from Sec.\ref{sec3}. Firstly, note that for a general bulk scalar potential $V(\Phi)$ and defining $z=r_c\phi$, the general scalar solution is given by
\begin{equation}\label{67}
    \Phi(z) = \Phi_{BR}(z)+\Phi_{OR}(z) = -\frac{k^{3/2}\alpha}{4}e^{4k(z-\pi r_c)} + \Phi_{OR} (z)
\end{equation}
Using the OR equation, we can now write
\begin{equation*}
\begin{split}
    \partial_z\Phi|_{\pi r_c} = -k^{5/2}\alpha - \frac{V'(\Phi_{OR})|_{\pi r_c}}{4kr_c}\\
    \partial_z V(\Phi)|_{\pi r_c} = V'(\Phi)|_{\pi r_c}(-k^{5/2}\alpha - \frac{V'(\Phi_{OR})|_{\pi r_c}}{4kr_c})
\end{split}
\end{equation*}
We write $V'(\Phi)|_{\pi r_c}$ as $V'(\Phi)$ and $V'(\Phi_{OR})|_{\pi r_c}$ as $V'(\Phi_{OR})$ for brevity. Then, plugging these along with the brane potentials defined in Sec.\ref{sec3} into Eq.(\ref{19}), applying Leibniz's rule to get the first derivative of the radion potential $V_{GW}(r_c)$ and setting it to zero (we demand the existence of an extremum at $r_c^*$) gives us the following relation evaluated at $r_c^*$
\begin{equation}\label{68}
\begin{split}
    \frac{\pi}{2}k^5\alpha^2 + \frac{\pi (V'(\Phi_{OR}))^2}{32k^2r_c^{*2}} + \frac{\pi k^{3/2}\alpha V'(\Phi_{OR})}{4r_c^*}-\pi k^{5/2}\alpha V'(\Phi) - \frac{\pi V'(\Phi)V'(\Phi_{OR})}{4kr_c^*}\\ + 2\pi \alpha^2 k^5 - 8\alpha \pi k^{7/2} \Phi_{OR}(\pi r_c^*) + 2\alpha k^{5/2}\frac{d\Phi_{OR}(\pi r_c)}{dr_c}|_{r_c^*} = 0
\end{split}
\end{equation}
$\Phi_{OR}(\pi r_c)$ is small as the outer region solution is suppressed at the boundary and $V'(\Phi)$ is assumed to be small at the boundary as per the CMS scheme. Thus, the terms $V'(\Phi)$ and $V'(\Phi_{OR})$ are small and we ignore their quadratics. This reduces Eq.(\ref{68}) to
\begin{equation}\label{69}
    \frac{5}{2}\alpha k^{7/2} + \frac{V'(\Phi_{OR})}{4r_c^*}-kV'(\Phi)-8k^2 \Phi_{OR}(\pi r_c) + 2k \Phi_{OR}'(\pi r_c) =0
\end{equation}
We now evaluate $V''_{GW}(r_c)$ at $r_c^*$.
\begin{equation}\label{70}
\begin{split}
    V''_{GW}(r_c)|_{r_c^*} = e^{-4kr_c^*\pi}[\frac{V''(\Phi_{OR})}{4r_c^*}.\frac{d\Phi_{OR}(\pi r_c)}{dr_c}-kV''(\Phi)\frac{d\Phi(\pi r_c)}{dr_c}\\-8k^2\frac{d\Phi_{OR}(\pi r_c)}{dr_c}+\frac{2k}{\pi}\frac{d^2\Phi_{OR}(\pi r_c)}{dr_c^2}-\frac{1}{4r_c^{*2}}V'(\Phi_{OR})]
\end{split}
\end{equation}
It should be remembered that the derivatives in the above expression are all evaluated at $r_c^*$. Also, from Eq.(\ref{67}), we see that $\frac{d\Phi(\pi r_c)}{dr_c} = \frac{d\Phi_{OR}(\pi r_c)}{dr_c}$. To ensure positivity of the radion mass and modulus stabilization, Eq.(\ref{70}) must be positive which results in the following criterion
\begin{equation}\label{71}
\begin{split}
    \frac{2k}{\pi}\frac{d^2\Phi_{OR}(\pi r_c)}{ dr_c^2}|_{r_c^*} > [8k^2+kV''(\Phi)|_{r_c^*}-\frac{V''(\Phi_{OR})|_{r_c^*}}{4r_c^*}]\frac{d\Phi_{OR}(\pi r_c)}{dr_c}|_{r_c^*}+ \frac{1}{4r_c^{*2}}V'(\Phi_{OR})|_{r_c^*}
\end{split}
\end{equation}
Using Eq.(\ref{69}) to substitute $\frac{d\Phi_{OR}(\pi r_c)}{dr_c}|_{r_c^*}$ in Eq.(\ref{71}), we get the expression
\begin{equation}\label{72}
\begin{split}
    (\frac{2k}{\pi})^2 \frac{d^2\Phi_{OR}(\pi r_c)}{ dr_c^2}|_{r_c^*} > [8k^2+kV''(\Phi)-\frac{V''(\Phi_{OR})}{4r_c^*}]\\ [kV'(\Phi) + 8k^2\Phi_{OR}(\pi r_c^*) - \frac{V'(\Phi_{OR})}{4r_c^*}-\frac{5}{2}\alpha k^{7/2}] + \frac{k}{2\pi r_c^{*2}}V'(\Phi_{OR})
\end{split}
\end{equation}
where on the right hand side, the \enquote{evaluated at $r_c^*$} is implicit. As one can see, the Eqs.(\ref{69}) and (\ref{72}) are completely determined by the bulk scalar potential (note that $\Phi(z)$ depends only on $V'(\Phi)$). Thus, these are the two relations that any general bulk potential must satisfy in order for the radion potential to have a stable minimum. It should again be noted that these results hold under the imposed boundary conditions.\\

\noindent We did a quick check for the case with no bulk interaction i.e. $V(\Phi) = \frac{1}{2}m^2\Phi^2$ with a set of parameters for which modulus stabilization is possible (see Eq.(2.22), \cite{Chacko1}): $\alpha=0.2, k=1, v=0.05, \epsilon = \frac{m^2}{4k^2} = 0.5$. This satisfies Eqs.(\ref{69}) and (\ref{72}) with $r_c^* = 1.156$. Similarly, one can do the checking for other potentials but these relations are guaranteed to hold as they are derived generally.\\

\noindent The radion mass squared is given by Eq. (\ref{70}) times $\frac{e^{2k\pi r_c^*}}{(Ak\pi)^2}$, since the canonical radion field is defined as $\varphi\equiv Ae^{-k\pi r(x)}$, where $A=\sqrt{24M^3/k}$. Thus, an overall warping of $e^{-k\pi r_c^*}$ emerges for the radion mass and it is expected to be $\mathcal{O}(TeV)$ in agreement with the existing literature \cite{GW2,csaki2000cosmology,csaki2001}. If one focuses on the positive terms in Eq. (6.4), then we can say that if a bulk potential is steep in the sense that $V''(\Phi_{OR})$ is large, and admits a rapidly varying analytic solution $\Phi_{OR}(\pi r_c)$ with $r_c$, atleast at $r_c^*$, then it ensures a heavy radion. The fields living on the visible brane couple to the radion through the induced metric \cite{kofman2004exact,csaki2001}, with an interaction $\frac{\varphi}{\sqrt{6}\Lambda_W} T^\mu_\mu$ where $\Lambda_W = M_{Pl} e^{-kr_c^*} \sim O({\rm TeV})$, $M^2_{Pl}= (1-e^{-2kr_c^*})/ (k \kappa^2)\sim 1/ (k \kappa^2)$, $T^\mu_\mu$ is the trace of the physical energy-momentum tensor of the TeV brane fields and $\kappa^2$ is the 5D Newton's constant. It is then clear that the radion couples as $\sim 1/$TeV to the Standard Model fields. An acceptable phenomenology requires $\mathcal{O}(TeV)$ radion mass which, as we saw, is likely in this context. As the dependence of the coupling on bulk potential is through the value of the stabilized modulus, it can be understood by analyzing Eq. (\ref{69}) for a given bulk potential, but a general feature is difficult to infer. Bounds on the radion mass and its coupling to visible sector fields can be found in Ref. \cite{bae2001bounds}.

\section{Stabilized modulus from brane world sum rules?}\label{sec7}
\noindent An essential feature one may have noticed while reading the paper till this point is that the analytic (or even numeric) calculation of the stabilized radion (if it gets stabilized) in each of the cases has been quite laborious. This in fact can be seen in several papers dealing with radion physics. So, is there an alternative way by which we could arrive at the stabilized modulus value - even in some cases? This is the question we want to address in this section.\\

\noindent Gibbons, Kallosh and Linde \cite{Gibbons} put down a set of consistency conditions for several braneworld scenarios, derived purely from the Einstein equations. These are widely known as the brane world sum rules or the GKL sum rules (after the authors). An extension of these rules in arbitrary dimensions was later provided by Leblond et al. \cite{Leblond}.\footnote{We refer the reader to Ref. \cite{Leblond} for a quick and self-contained review of the brane world sum rules - we do not go over the derivation in this paper.} In this section, we show for the first time, that in certain Goldberger-Wise scenarios where stabilization occurs, the GKL rules can potentially help pin down the value of the stabilized modulus. This holds true even when gravitational contribution to the radion potential is included.\\

\noindent One of the GKL conditions is the check that the summation of the flat brane tensions and the non-negative integral of the gradient energy of the bulk scalars must vanish. The original GW scheme did not respect this criterion and hence was the need to involve backreaction and solve for the combined gravity-scalar system. However, in generalized GW scenarios (see Sec.3), the choice of a linear visible brane potential and the fine-tuning of the 4D cosmological constant to be negligible lets us use the brane sum rules to get to a non-trivial inequality expression involving the stabilized modulus ($r_c$) and the parameters in our setup.\footnote{In both the cases discussed below, the parameters $\alpha$ and $v$ are taken to be small and almost of the same order.}\\

\noindent The brane world sum rule of interest to us is thus given by (see Section 2, \cite{Gibbons})
\begin{equation}\label{88}
     \sum_{i=\{vis, hid\}} \lambda_i(\Phi_i)  + \oint
\Phi'\cdot \Phi'  =0 \ .
\end{equation}
where $\lambda_i (\Phi_i)$ is the flat brane tension (including the brane potential for the scalar field $\Phi$) on the corresponding (visible or hidden) brane, and the second term corresponds to the non-negative gradient energy of the scalar. As a result, the brane tensions must obey
\begin{equation}\label{89}
    \sum_{i=\{vis, hid\}} \lambda_i(\Phi_i) \leq 0
\end{equation}

\noindent As per our setting for the generalized GW scenarios, the \enquote{$\lambda_i$}s take the form
\begin{equation}\label{90}
    \lambda_{hid} = \gamma (\Phi(0)^2 - k^3v^2)^2 + T_{hid}
\end{equation}
\begin{equation}\label{91}
    \lambda_{vis} = 2k^{5/2}\alpha\:\Phi(\pi) + T_{vis}
\end{equation}
where $T_{hid}$ and $T_{vis}$ are the brane tensions in absence of the scalar field. Recalling $\Lambda_{4D} = (T_{hid}+\frac{\Lambda_b}{k})/k^4$ to be the 4D cosmological constant which we tuned to be small as in Ref. \cite{Chacko1}, $T_{hid} \approx -\Lambda_b/k$. Therefore, $\tau = (T_{vis} + T_{hid})/k^4$. A thing to notice is that in the original GW scheme, $\Phi (0)$ was not exactly equal to $k^{3/2}v$. Rather, they assumed a large coupling constant ($\gamma$) using which one could show that $\gamma (\Phi(0)^2 - k^3v^2)^2 = \frac{m^4}{64k^2 \gamma}$. This vanished in the limit $\gamma \rightarrow \infty$. However, in our present setup, we assume the boundary condition $\Phi(0) = k^{3/2} v$ which straightaway sets the first term in Eq.(\ref{90}) to zero.\\

\noindent It might be confusing to see that $(T_{hid} + T_{vis})$ does not vanish. This comes because we promote the flat 4D background to a general metric. As a general metric no longer requires the fine-tuning in Eq. (\ref{6}), the $(T_{hid} + T_{vis})$ sum is non-zero and thereby ensures a non-zero gravitational contribution to the radion potential (see Sec.\ref{sec2.2}). When we do not replace $\eta_{\mu\nu}$ with $g_{\mu\nu}$, the fine-tuning remains intact and the gravitational contribution vanishes as has mostly been the case in this paper till now.\\

\noindent Adding Eqs.(\ref{90}) and (\ref{91}), we get
\begin{equation}\label{92}
    \lambda_{hid} + \lambda_{vis} = 2k^{5/2}\alpha\:\Phi(\pi) + k^4 \tau
\end{equation}
Finally, plugging this result into Eq.(\ref{89}) gives us
\begin{equation}\label{93}
    2\alpha\:\Phi(\pi) + k^{3/2} \tau \leq 0
\end{equation}
In what follows, we will work with the above criterion which serves as a tailored brane sum rule for such generalized GW scenarios. The only unknown in Eq.(\ref{93}) which is determined by our choice of the bulk scalar potential is the value of the scalar field at the visible brane i.e. $\Phi (\pi)$. Hence, the only step needed to get the inequality expression for the stabilized modulus is to solve for the scalar field profile.\\

\noindent As discussed previously, a class of bulk scalar potentials where radion stabilization is guaranteed is where they possess a global minimum. Thus, we consider a scalar field where the bulk potential is dominated by the mass term, 
\begin{equation}\label{94}
    V_b(\Phi)=\frac 12m^2\Phi^2\;
\end{equation}
Following the steps as in Sec.3, one then obtains the following solution for the scalar field profile
\begin{equation}\label{95}
    \Phi(\phi)=-\frac{k^{3/2}\alpha}
{4}e^{4kr_c(\phi-\pi)}+k^{3/2}ve^{-\epsilon kr_c\phi}\;
\end{equation}
where $\epsilon = m^2/4k^2$ with $|\epsilon| \ll 1$ (as $\alpha v > 0$, $m^2$ has to be positive to be consistent with the expression for mass of the radion \cite{Chacko1}). Now, following the steps as discussed in Sec.\ref{sec3.2} allows one to get to the expression for the radion potential up to the leading order in $v$ and $\epsilon$, including the gravitational contributions \cite{Chacko1}.
\begin{equation}\label{96}
V(\varphi) =
2k^4\alpha v\left(\frac{\varphi}{A}\right)^{4+\epsilon}
+k^4 \tau \left(\frac{\varphi}{A}\right)^4\;
\end{equation}
where $\tau$ is redefined as $\tau \coloneq (\tau -\frac{\alpha^2}{4})$. The minimum of this potential occurs for the condition:
\begin{equation}\label{97}
    e^{-\epsilon kr_c\pi} = -\frac{\tau}{2\alpha v}+\mathcal{O}(\epsilon)
\end{equation}
Thus, we see that Eq.(\ref{97}) gives us an expression for the stabilized modulus in this setup. It should be noted that $\alpha v>0$ enforces $\tau$ to be negative, the left side of the equation being an exponential. Now, lets turn our attention to the brane world sum rules and see if we could get somewhere close to this result. Plugging in the solution $\Phi (\pi)$ from Eq.(\ref{95}) into the tailored brane sum rule for such cases (Eq.\ref{93}) gives us
\begin{equation}\label{98}
    \tau - \frac{\alpha^2}{4} + 2\alpha v e^{-\epsilon kr_c \pi} \leq 0
\end{equation}
This leads us to the following inequality expression for the stabilized modulus ($r_c$),
\begin{equation}\label{99}
    e^{-\epsilon kr_c \pi} \leq -\frac{\tau}{2\alpha v} + \frac{\alpha}{8v}
\end{equation}
As both $\alpha$ and $v$ are small parameters and almost of the same order, we can drop the second term in Eq.(\ref{99}) in favour of the first term which is indeed dominating. This allows us to get to the result
\begin{equation}\label{100}
    e^{-\epsilon kr_c \pi} \leq -\frac{\tau}{2\alpha v}
\end{equation}
Thus, we get an effective lower bound on $r_c$ and we notice that this very well coincides with the true value of the stabilized modulus. If the gravitational part were zero, then the right side of Eq. (\ref{99}) becomes $\frac{\alpha}{4v}$. As the stabilized radion value from Eq. (\ref{97}) would then be $\frac{\alpha}{8v}$, we see that the non-trivial bound gives the correct order of magnitude.\\

% \noindent As $\alpha v > 0$, we get yet another condition on the parameters involved from Eq.(\ref{98}). Dividing the whole expression by $2 \alpha v$, we see that the last term is non-negative being an exponential. Hence, the other two terms have to add up to negative yielding $\alpha^2 > 4 \tau$.
% This is a weaker condition than $\tau$ being negative. However, the fact that $\alpha$ is small reduces this condition to approximately $\tau < 0$ (we also see this from Eq.\ref{100}), which is consistent with our prior finding.

\noindent Can the brane-sum bounds be helpful in other scenarios? - say for the bulk cubic interaction. This calculation was already performed in Sec.\ref{sec4.2} (including gravitational contributions, the $-\frac{\alpha^2}{4}$ in Eq.(\ref{33}) and further equations should be replaced by $\tau$ - the approximate stabilized modulus value will not change). Thus, plugging in the solution of the scalar field from Eq.(\ref{31}) into Eq.(\ref{93}) gives us 

\begin{equation}\label{101}
    \frac{4\alpha v}{1+\xi kr_c \pi} + 2\tau - \alpha^2 \leq 0
\end{equation}
Assuming $\frac{4\alpha v}{1+\xi kr_c \pi}$ and $\alpha^2 - 2\tau$ have the same sign \footnote{If they were taken to be of opposite signs, the nature of bound in Eq.(\ref{102}) and consequently Eq.(\ref{103}) would be flipped i.e. a lower bound on $r_c$.}, we get the following bounding criterion for the stabilized modulus,
\begin{equation}\label{102}
    k\pi r_c \leq -\frac{1}{\xi} + \frac{4\alpha v}{\xi (\alpha^2 - 2\tau)}
\end{equation}
As $\xi \small \propto v$, the second term is independent of $v$ and hence, is proportional to $\frac{1}{\eta(\alpha - 2\tau/\alpha)}$. Now, $\alpha$ being a small parameter and $\eta$ being not so small (as the cubic term dominates the bulk potential) make the denominator very large, thus making the second term contribute negligibly compared to the first one. Therefore,
\begin{equation}\label{103}
    k \pi r_c \leq -\frac{1}{\xi}
\end{equation}
Hence, we get an upper bound on $r_c$ which due to its proximity to the true value (Eq.\ref{35}) is yet again, a great estimate for the stabilized modulus.

\section{Discussion}\label{sec8}
\noindent The study brings out a deep correlation between the form of the bulk scalar potential - more generally, the scalar field profile (which has been employed for modulus stabilization) and the corresponding existence of a stable minimum for the radion potential. Our calculations addressed various scalar field actions to explore the resulting stabilization of the moduli sector. This included a canonical kinetic term as well as phantom like term for the scalar sector. Moreover, choosing different forms of the scalar field potentials, we carefully examined the possibility of a corresponding modulus stabilization. A summary of these cases can be found in Sec.\ref{sec4.7}. We also demonstrated that the qualitative nature of bulk potentials can hint at the fate of modulus stabilization. In particular, the existence of a global minimum for the scalar sector was shown to ensure a stable minimum for the modulus. More precisely, the existence of a global minimum in the bulk potential is sufficient but not a necessary condition for modulus stabilization. On the contrary, the scalar sector which does not have a global minimum may or may not lead to radion stabilization. Taking into account the stability of the bulk scalar, this situation gets further classified: a stable bulk scalar always stabilizes the modulus, and failing to stabilize the radion can only result from an unstable bulk scalar. We should remember that all these conclusions hold provided an \textit{appropriate} extremum for the radion potential exists in the first place. We then deduced the necessary criteria that any bulk scalar potential must satisfy to enable radion stabilization. It must be noted that our analysis consistently employed the CMS boundary conditions. The physical motivation behind choosing these conditions was explained in Sec.\ref{sec3.1}. In principle, one is free to explore innumerable boundary conditions - each of these may have its own physical justification and they all serve as potential future works. The only instance where we deviated from CMS boundary conditions was when we discussed a non-trivial backreacted geometry with the BFG potentials. In this regard, we should mention that even under the original Goldberger-Wise boundary conditions, a global minimum in bulk potential stabilized the radion, whereas potentials with a global maximum and no global minimum (both the stable and unstable bulk scalars) failed to stabilize the radion. It may further be observed that such different stabilizing scalar sectors in the scalar-tensor models have direct correspondence with higher curvature gravity models and thus, our calculations indirectly include various viable $f(R)$ models of gravity which may lead to modulus stabilization \cite{banerjeepaul,shafaq1}.\\

\noindent Regarding the CMS boundary conditions, we already highlighted their physical motivation, but one must also remember that they are most naturally motivated by our singular perturbation theory-based solution technique. If one sees our outer region (OR) and boundary region (BR) equations (see Sec.\ref{sec3.1}), we certainly need a Dirichlet boundary condition at the hidden brane since there is only a first order derivative of $\Phi$ in the OR equation, whereas the BR equation demands a Neumann boundary condition at the visible brane (since there is a second order derivative of $\Phi$ in the BR equation). This is ensured by a linear visible brane potential. Hence, we are naturally led to the CMS boundary conditions to make the solution technique work. If one wishes to work with the original GW Dirichlet boundary conditions on both branes, then the present analytic solution technique and they have to resort to numerics. Even then, they have to check large parameter spaces before reaching any general result.\\

\noindent While the five-dimensional Randall--Sundrum (RS1) model provides an elegant resolution to the gauge hierarchy problem, it faces increasing tension with present collider constraints. In particular, the non-observation of an $\mathcal{O}(\text{TeV})$ Kaluza--Klein graviton mode at the ATLAS and CMS experiments has progressively restricted the viable parameter space of the model \cite{CMS:2024nht}. To remain consistent with these bounds, one must now invoke a modest hierarchy of order $\lesssim 10^{-2}$ between the five-dimensional Higgs vacuum expectation value and the five-dimensional Planck scale---an assumption that introduces its own theoretical challenges. In particular, the fine-tuning problem appears in a new form and spoils the elegance of the original RS1 model, whose success was to resolve the hierarchy problem of the Higgs mass without fine-tuning of the fundamental parameters (see, e.g., the introductory discussions of \cite{arun2015graviton,bhaumik2024moduli}). A natural way to circumvent these issues is to consider higher-dimensional generalizations of the RS framework with multiple warping, which have been extensively studied in Refs. \cite{livingedge,mukhopadhyaya2012matter,chakraborty2014bulk,arun2015graviton,bhaumik2022moduli,bhaumik2023nested,bhaumik2024moduli}. These multiply warped models can remain compatible with current collider bounds while continuing to address the hierarchy problem effectively. It would therefore be worthwhile to extend and apply the present methodology to such higher-dimensional setups in future works.\\

\noindent This paper also discusses a potential connection between geometric consistency conditions and modulus stabilization. We showed that in cases where the bulk scalar potential possesses a global minimum and thereby stabilizes the radion, the GKL sum rules can help us pin down the value of the stabilized radion. This may seem unexpected at first since getting to the stabilized modulus is different physics altogether. However, we could argue with a possible explanation - since there are an infinite number of GKL rules (they form a one-parameter family of conditions), these constraints essentially reduce the solution space of the modulus to a point that the only self-consistent solution is the stabilized one (subject to future exploration) - but still, why so? - we do not have an answer yet. Also, in the specific cases that we discussed, the fact that the visible brane potential did not vanish when evaluated at $\phi = \pi$ did the main trick as it gave away the radion contribution when plugged into the sum rule. Exploring other boundary conditions and brane potentials with such properties is thus quite lucrative. We also demonstrated that upon bulk cubic interaction, the sum rules offer the same insight. Even if one is hesitant to view this as an exact determination, it still serves as a great order estimation protocol. Thus, though this paper provides the first steps, whether the connection between brane sum rules and modulus stabilization goes far beyond the cases that we discussed, and whether we can understand any possible underlying physics are subject to further research. Some of these will be addressed in our future works.\\

\section*{Acknowledgements}
\noindent SB acknowledges helpful discussions with Prof. Sourov Roy and Prof. Dilip Kumar Ghosh. We thank both the anonymous referees for their insightful comments which have led to the enhancement of this manuscript. SB is supported through INSPIRE-SHE Scholarship by the Department of Science and Technology (DST), Government of India.

\section*{Data Availability Statement}
Being a theoretical study, no experimental data is associated with this work.

\section*{Code Availability Statement}
Apart from numerical integration, no major code is associated with this work. This shall be made available on reasonable request.

\appendix

\section{Including backreaction - the superpotential method}\label{appa}
\noindent In GW scenarios, we neglected the {\it back-reaction} of the metric to the presence of the scalar
field in the bulk but this is important indeed and therefore, it would be very nice to simultaneously solve the Einstein and the
bulk scalar equations, to have the back-reaction exactly under control. DeWolfe et al. \cite{Freedman} provides us with such a formalism.
Denote the scalar field in the bulk by $\Phi$, and consider the action
\begin{equation}\label{a1}
\begin{split}
\int d^5x\sqrt{g} \left[-M^3 R++\frac{1}{2}(\nabla\Phi)^2 -V(\Phi )\right]-
\int d^4x \sqrt{g_4} \lambda_{h} (\Phi )-
\int d^4x \sqrt{g_4} \lambda_{v} (\Phi )
\end{split}
\end{equation}
We look for an ansatz
of the background metric again of the generic form as in the RS case to maintain 4D Lorentz invariance:
\begin{equation}\label{a2}
ds^2 =e^{-2 A(y)} \eta_{\mu\nu} dx^\mu dx^\nu -dy^2
\end{equation}
Using this metric, we get the Einstein equations and bulk scalar equation of the form
\begin{align}
4 A'^2 - A'' &= -\frac{2\kappa^2}{3} V(\Phi_0) - \frac{\kappa^2}{3} \sum_{i=vis,hid} \lambda_i (\Phi_0) \delta (y-y_i), \label{a3} \\
A'^2 &= \frac{\kappa^2}{12} \Phi_0'^2 - \frac{\kappa^2}{6} V(\Phi_0), \label{a4} \\
\Phi_0'' - 4A'\Phi_0' &= \frac{\partial V}{\partial\Phi_0} + \sum_i \frac{\partial\lambda_i (\Phi_0)}{\partial\Phi} \delta (y-y_i). \label{a5}
\end{align}

\noindent The jump conditions at the branes are then given by
\begin{eqnarray}\label{a6}
&& [A']_i=\frac{\kappa^2}{3}\lambda_i(\Phi_0) \nonumber \\
&& [\Phi_0']_i=\frac{\partial\lambda_i (\Phi_0)}{\partial\Phi}
\end{eqnarray}
The bulk equations Eqs. (\ref{a3}-\ref{a5}) along with the boundary conditions Eq. (\ref{a6}) form the coupled gravity-scalar system. We can now define the function $W(\Phi )$ via the equations
\begin{eqnarray}\label{a7}
&&A'\equiv \frac{\kappa^2}{6} W (\Phi_0) \nonumber \\
&&\Phi_0'\equiv \frac{1}{2} \frac{\partial W}{\partial \Phi}
\end{eqnarray}
Plugging in these expressions into the bulk equations, we find that all the equations are satisfied simultaneously if the following consistency criterion holds,
\begin{equation}\label{a8}
V (\Phi )=\frac{1}{8} \left( \frac{\partial W}{\partial \Phi}\right)^2-\frac{\kappa^2}{6} W(\Phi )^2
\end{equation}
The jump conditions translate to:
\begin{eqnarray}\label{a9}
&& \frac{1}{2}[W(\Phi_0)]_i=\lambda_i(\Phi_0) \nonumber \\
&& \frac{1}{2}[\frac{\partial W}{\partial\Phi}]_i=\frac{\partial\lambda_i(\Phi_0)}{\partial\Phi}
\end{eqnarray}
These jump conditions are satisfied only when the brane potentials are chosen of the form
\begin{equation}\label{a10}
\lambda_h = W(\Phi_h)+ W'(\Phi_h) (\Phi-\Phi_h)+
\gamma_h (\Phi-\Phi_h)^2
\end{equation}
and
\begin{equation}\label{a11}
\lambda_v = -W(\Phi_v)- W'(\Phi_v) (\Phi-\Phi_v)+
\gamma_v (\Phi-\Phi_v)^2
\end{equation}

\noindent Thus, we see that that the coupled second order differential equations now reduce to ordinary first order equations. The difficulty one has to entail is to find the superpotential $W(\Phi)$ given the potential $V(\Phi)$ which generally is a hard task. However, if we only need a superpotential which produces a family of potentials with some very general properties, then this prescription certainly simplifies our working.

\section{A relation between $r_c$ and p for the BFG case}\label{appb}
\noindent Given that we take our conjecture holds true that there is indeed modulus stabilization for all BFG type potentials, a natural question to ask is whether we can establish some relation between $r_c$ and the only parameter in the problem p. We listed down all known $r_c$ values till p = 19, and found the empirical expression which best-fits the $r_c$ vs p curve as
\begin{equation}\label{b1}
    \pi r_c = ap^{1/b}+c
\end{equation}

% \begin{figure}[h]\label{f9}
%     \centering
%     \includegraphics[width=0.95\linewidth]{r_vs_p_total.png}
%     \caption{List of known $r_c$ values (left) and plot for $\pi r_c$ vs p (right) for p till 19}
%     \label{fig:enter-label}
% \end{figure}\\

\noindent with a = 1.16, b = 0.82 and c = -0.69.

% \noindent To put the formula to test, we let our numerics run for quite some time and came up with the numerical values of $\pi r_c$ for p = 35, 37 and 39. The respective true values are 87.35, 92.77 and 100 while Eq.(\ref{b1}) gives us the values as 87.47, 93.64 and 99.89.

\bibliographystyle{JHEP}

\bibliography{main}

\end{document}